\documentclass[prb,draft,amsfonts,amssymb,amsmath,showpacs]{revtex4}
\begin{document}

\title{Dynamical Symmetry Breaking of a Relativistic Model in Quasi-(1+1)-Dimensions. I. Formulation}
\author{Tadafumi Ohsaku}
\affiliation{Institut f\"{u}r Theoretische Physik, Universit\"{a}t zu K\"{o}ln, 50937 K\"{o}ln, Deutschland}
\date{\today}

\newcommand{\bmx}{\mbox{\boldmath $x$}}
\newcommand{\bmp}{\mbox{\boldmath $p$}}
\newcommand{\bmk}{\mbox{\boldmath $k$}}
\newcommand{\bmq}{\mbox{\boldmath $q$}}
\newcommand{\Afey}{\ooalign{\hfil/\hfil\crcr$A$}}
\newcommand{\kfey}{\ooalign{\hfil/\hfil\crcr$k$}}
\newcommand{\pfey}{\ooalign{\hfil/\hfil\crcr$p$}}
\newcommand{\partfey}{\ooalign{\hfil/\hfil\crcr$\partial$}}
\newcommand{\Dfey}{\ooalign{\hfil/\hfil\crcr$D$}}
\newcommand{\hfey}{\ooalign{\hfil/\hfil\crcr$h$}}

\begin{abstract}

The dynamical symmetry breaking in a quasi-(1+1)-dimensional relativistic model is investigated.
The motions of particles in intrachain are described as a relativistic electron-hole gas, 
while the interchain hopping term is introduced as a 0th-component of vector in (1+1)-dimensions, a kind of chemical potential of the system.
The gauge symmetry of the model is chosen as $U(1)$ suitable for a possible situation of a real substance in condensed matter physics.
We consider the BCS-type contact interactions for the $s$-wave fermion-pair condensates, 
while employ the nonlocal interactions of the generalized BCS framework to generate the $p$-, $d$- and $f$-wave condensations in the system. 
Especially we examine the dynamical generation of a Dirac mass term and superconductivity in the model.
The phenomenon is interpreted as metal-insulator/metal-superconductor phase transitions.

\end{abstract}


\maketitle

\section{Introduction}

Recently, many papers have considered the competitions/coexistences of several orders coming from particle-hole and particle-particle type fermion pair condensations~[1-11].
For example, in quasi-(1+1)-dimensional organic superconductors ( the Bechgaard salts $({\rm TMTTF})_{2}X$ )~[3-6,8-11], 
the competition between instabilities toward taking place of superconducting states ( $d$-wave singlet or $f$-wave triplet ) 
and spin/charge density waves ( SDW/CDW ) was discussed in the model parameter space by the renormalization group analysis~[9-11]. 
Another interesting coexistence of the diagonal and off-diagonal long range orders might be found in the physics of supersolid~[12].
In this case, the possibility of the coexistence of the diagonal ( CDW ) and off-diagonal ( superfluidity ) long range orders are discussed.
These orders seem to have a universal relation under variations with respect to external and/or model parameters like chemical potential, temperature, coupling constant for interaction between particles:
Various substances shows superconductivity, and in their phase diagrams, there are several orderings, CDW, SDW, Mott-Hubbard insulator, so forth, in the neighbor of the superconducting states.
The competition between fermion-antifermion ( for example, SDW, CDW, mesons, dynamical chiral symmetry breaking ) and fermion-fermion ( diquarks, superconductivity ) condensations are observed/considered under various situations in interacting Fermi systems, in condensed matters physics, nuclear many-body systems, quark matters~[13,14], so forth.
This feature exists both in the description of phenomenological BCS ( Bardeen, Cooper, Schrieffer ) theories~[15-20] and theories on mechanisms of the origin of superconductivity/superfluidity~[2,8-11],
because the competition of these orders has the origin in the different types of excitations of interacting Fermi systems, particle-hole or particle-particle types.
Quite a large part of physical properties of interacting Fermi systems of condensed matters will be determined by the existence of the Fermi surface or the gap ( mass ) generation at the Fermi level.
There are mainly three mechanisms of the mass generation: the Higgs, the BCS-NJL ( Nambu and Jona-Lasinio )~[21] and the Kaluza-Klein mechanisms.
Especially in the situation of interacting Fermi systems of condensed matters, the BCS-NJL mechanism is the suitable method to describe the physical property of the systems.
In this paper, we examine physical properties of a quasi-(1+1)-dimensional relativistic model under competition/coexistence of several types of chiral condensations and BCS and generalized-BCS superconductivities~[22-26].

This paper is organized as follows. In Sec. II, we introduce our model Lagrangian, discuss its structure and symmetries.
In Sec. III, the group theoretical consideration of the order parameters they are examined in this work is presented.
The effective potential and the gap equations are obtained in Sec.IV.
For generating not only $s$-wave condensations but also $p$-, $d$- and $f$- wave pairings of both fermion-antifermion and fermion-fermion types,
we prepare the generalized BCS formalism.
Finally in Sec. V, we give the summary of this work.
The numerical part will be published as the part II of this study.

\section{The Quasi-(1+1)-Dimensional Relativistic Model}

We consider quasi-one-dimensional electron system~[3,4]. 
The system has three spacetime coordinates $x=(x_{0},x_{1},x_{2})$.
One-dimensional $L$ chains are stacked in parallel with a fixed periodicity.
The thermodynamic limit is taken by $L\to \infty$ in this case.
$x_{2}$-coordinate is discretized, $x_{2}=n_{x}\in{\bf N}$, $1\le n_{x}\le L$, and denotes the site $n_{x}$.
Particles ( electrons and holes ) behave as a homogeneous Fermi gas toward $x_{1}$ ( intrachain ) direction, 
while they move to $x_{2}$-direction by the interchain nearest neighbor hopping with a parameter $t$ of the tight-binding picture~[27]. 
$t$ ( $\in {\bf R}$ ) corresponds to the overlapping integrals of atomic orbitals between two nearest chains. 
In the vicinity of the Fermi energy, 
energy-momentum relation will be approximately described by the linear-dispersion relation $\pm v_{F}|k_{1}|$ ( $v_{F}$; the (1+1)-dimensional Fermi velocity ).
The energy spectra of particles in such a system are given as follows~[4,9-11]:
\begin{eqnarray}
{\cal E}_{\pm}(\bmk) &=& \pm v_{F}|k_{1}| -2t\cos k_{2}.
\end{eqnarray}
Here, $\bmk=(k_{1},k_{2})$. 
We will examine the phenomenon of dynamical symmetry breaking in the quasi-one-dimensional electron system.
For this purpose, we employ the following Lagrangian density of a Dirac model: 
\begin{eqnarray}
{\cal L} &=& {\cal L}_{em} + {\cal L}_{0} + {\cal L}^{s}_{I} + {\cal L}^{nl-d}_{I} + {\cal L}^{nl-od}_{I} +{\cal L}_{B},  \\
{\cal L}_{em} &\equiv& -\frac{1}{4}F_{\mu\nu}F^{\mu\nu}, \\
{\cal L}_{0} &\equiv&  \bar{\psi}_{\alpha}(x_{0},x_{1},n_{x})\Bigl[ \gamma^{0}\{ iD_{0} + v_{F}k_{F}-t \} +iv_{F}\gamma^{1}D_{1} -u \Bigr] e^{ie\widetilde{A}_{2}(n_{x}+1,n_{x})}\psi_{\alpha}(x_{0},x_{1},n_{x}+1)  \nonumber \\ 
& & + \bar{\psi}_{\alpha}(x_{0},x_{1},n_{x}+1) \Bigl[ \gamma^{0}\{ iD_{0} + v_{F}k_{F}-t \} +iv_{F}\gamma^{1}D_{1} -u \Bigr] e^{-ie\widetilde{A}_{2}(n_{x}+1,n_{x})}\psi_{\alpha}(x_{0},x_{1},n_{x}),   \\
{\cal L}^{s}_{I} &\equiv& \frac{G_{0}}{2N}\Bigl[ (\bar{\psi}_{\alpha}(x)\psi_{\alpha}(x))^{2} + (\bar{\psi}_{\alpha}(x)i\gamma_{5}\psi_{\alpha}(x))^{2} \Bigr] = \frac{G_{0}}{2N}(\bar{\psi}_{\alpha}\gamma^{\mu}\psi_{\alpha})(\bar{\psi}_{\beta}\gamma_{\mu}\psi_{\beta}),  \\
{\cal L}^{nl-d}_{I} &\equiv& \frac{1}{2N} \sum_{n_{y},n_{x'},n_{y'}}\int dx'_{0}dx'_{1}dy_{0}dy_{1}dy'_{0}dy'_{1} \bar{\psi}_{\alpha}(x)\psi_{\alpha}(y)V(x,y;x',y')\bar{\psi}_{\beta}(x')\psi_{\beta}(y'),   \\
{\cal L}^{nl-od}_{I} &\equiv& \frac{1}{2N}\sum_{n_{y},n_{x'},n_{y'}}\int dx'_{0}dx'_{1}dy_{0}dy_{1}dy'_{0}dy'_{1}\bar{\psi}_{\alpha}(x)\bar{\psi}^{T}_{\alpha}(y)W(x,y;x',y')\psi^{T}_{\beta}(x')\psi_{\beta}(y'), \\
{\cal L}_{B} &\equiv& -g_{J}\mu_{B}\vec{B}\cdot\bar{\psi}_{\alpha}\gamma^{0}\frac{\vec{\tau}_{\alpha\beta}}{2}\psi_{\beta},  
\end{eqnarray}
where, the definitions of fields and convariant derivatives are given as follows:
\begin{eqnarray}
F_{\mu\nu} &\equiv& \partial_{\mu}A_{\nu}-\partial_{\nu}A_{\mu},  \quad
\psi_{\alpha} \equiv (R_{\alpha}(x),L_{\alpha}(x))^{T}, \quad \alpha,\beta = \uparrow, \downarrow, \quad
D_{0} \equiv \partial_{0} -ieA_{0}, \quad D_{1} \equiv \partial_{1} -ieA_{1}.
\end{eqnarray}
${\cal L}_{em}$ and ${\cal L}_{0}$ are the kinetic terms of electromagnetic and matter fields, respectively. 
${\cal L}^{s}_{I}$ is a four-fermion contact ( $s$-wave ) intrachain interaction between particles.
$G_{0}$ is a bare coupling constant with mass dimension $[{\rm mass}]^{-1}$. 
On the other hand, ${\cal L}^{nl-d}_{I}$ ( ${\cal L}^{nl-od}_{I}$ ) has the nonlocal interaction between particles $V$ ( $W$ ), 
can include both inter- and intra-chain interactions. 
It is prepared for generating $p$-, $d$- and $f$- wave fermion-antifermion ( fermion-fermion ) pair correlations.
$\psi_{\alpha}$ is a two-component Dirac field, has the (iso)spin index $\alpha$ as an isodoublet, 
replicated into an $N$-flavor field for taking the large-$N$ limit of $1/N$ expansion smoothly. 
$R_{\alpha}(x)$ ( $L_{\alpha}(x)$ ) describes particles move toward the right ( left ) direction.
We should note that the Dirac field $\psi_{\alpha}$ is defined in (2+1)-dimensional spacetime, 
thus the mass dimension of the field is determined through the dimensional analysis in (2+1)-dimensions ( not in (1+1)-dimensions ):
$\psi_{\alpha}$ carries the mass dimension $[{\rm mass}]^{1}$.
On the contrary, the definition of gamma matrices are chosen as the ordinary (1+1)-dimensional field theory~[28,29]: 
\begin{eqnarray}
\gamma^{0} &=& \sigma_{1} = \left(
\begin{array}{cc}
0 & 1 \\
1 & 0
\end{array}
\right), \quad 
\gamma^{1} = -i\sigma_{2} = \left(
\begin{array}{cc}
0 & -1 \\
1 & 0
\end{array}
\right), \quad \gamma_{5} = \gamma^{0}\gamma^{1} = \left(
\begin{array}{cc}
1 & 0 \\
0 & -1 
\end{array}
\right).
\end{eqnarray}
They satisfy the following Clifford algebra relations $\{\gamma^{\mu},\gamma^{\nu}\}=2g^{\mu\nu}$ 
( we will use the Lorentz signature $g^{\mu\nu}={\rm diag}(1,-1)$ throughout this paper. No euclidization will be needed ).
The charge conjugation matrix $C\equiv \gamma^{1}$ ( $C^{-1} = -\gamma^{1}$ ) satisfies $C\gamma^{\mu}C^{-1}=-\gamma^{\mu T}$.
$\gamma_{5}$ anticommutes with $C$ in (1+1)-dimensions. 
${\cal L}_{0}$ has couplings between matter $\psi_{\alpha}$, $\bar{\psi}_{\alpha}$ and $U(1)$ gauge fields $A_{\mu}$ ( $\mu=0,1,2,3$ ). 
The local $U(1)$ gauge invariance will be maintained under the following gauge transformation law~[28]:
\begin{eqnarray}
& & A_{\mu}(x) \to A_{\mu}(x) + \frac{1}{e}\partial_{\mu}\theta_{V}(x), \quad ( \mu = 0,1,2,3 ), \quad \psi_{\alpha}(x) \to e^{-i\theta_{V}(x)}\psi_{\alpha}(x),  \nonumber \\
& & \widetilde{A}_{2}(n_{x}+1,n_{x}) \equiv \int^{n_{x}+1}_{n_{x}}dsA_{2}(s),  \quad \widetilde{A}_{2}(n_{x}+1,n_{x}) \to \widetilde{A}_{2}(n_{x}+1,n_{x}) + \frac{1}{e}\Bigl( \theta_{V}(n_{x}+1) - \theta_{V}(n_{x})\Bigr).      
\end{eqnarray}
In the integral for the definition of $\widetilde{A}_{2}(n_{x}+1,n_{x})$, 
it should be understood that the integration parameter $s$ continuously runs from the coordinate of site $n_{x}$ to site $n_{x}+1$. 
The interaction between the charge density and the vector potential $A_{0}\bar{\psi}\gamma^{0}\psi$ is already included in ${\cal L}_{0}$.
${\cal L}_{B}$ is the Zeeman term which would give the Pauli spin polarization, $g_{J}$ is the $g$-factor, and $\mu_{B}$ is the Bohr magneton.
$\vec{B}=(B_{1},B_{2},B_{3})=-\epsilon_{ijk}\partial_{j}A_{k}$ ( $\epsilon_{123}=1$, $i,j,k=1,2,3$ ) are magnetic fields.
The spin polarization has been introduced as a 0th component of vector in (1+1)-dimensional Lorentz symmetry. 
Hence, the Zeeman term acts as different effective chemical potentials of different spin eigenstates.
In the definition of ${\cal L}$, we have chosen that the third component of $A_{\mu}$ can couple with matter only through the Zeeman term.
$\psi_{\alpha}$ and $\bar{\psi}_{\alpha}$ live in (2+1)-dimensional spacetime, 
while the magnetic field $\vec{B}$ has three components and defined in (3+1)-dimensions:
The gauge fields $A_{\mu}$ can propagate in (3+1)-dimensional spacetime.
Throughout this paper, it is enough for us to evaluate several quantities under the (2+1)-dimensional formalism.
Hereafter, $A_{\mu}$ in ${\cal L}_{0}$ are omitted, and we regard $\vec{B}$ as external fields ( not a dynamical degree of freedom ).
The non-vanishing $\vec{B}$ fields break the time-reversal invariance of the system.
In our system, the Fourier transform of $\psi_{\alpha}$ becomes
\begin{eqnarray}
\psi_{\alpha}(x_{0},x_{1},n_{x}) &=& \int_{k}\psi_{\alpha}(k_{0},k_{1},k_{2})e^{-ik_{0}x_{0}+ik_{1}x_{1}+ik_{2}n_{x}}, \\
\int_{k}(\cdots) &\equiv& \int\frac{dk_{0}}{2\pi}\int\frac{dk_{1}}{2\pi}\int^{\pi}_{-\pi}\frac{dk_{2}}{2\pi}(\cdots).
\end{eqnarray} 
The domain for the integration $-\pi\le k_{2}\le \pi$ corresponds to the first Brillouin zone in $k_{2}$ direction.
Hence, the kinetic term obtains the following Fourier transformed expression:
\begin{eqnarray}
{\cal L}_{0} &=& \bar{\psi}_{\alpha}\Bigl(\gamma^{0}\{ k_{0} + v_{F}k_{F}+\mu(k_{2}) \}+v_{F}\gamma^{1}k_{1} -m(k_{2})  \Bigr)\psi_{\alpha},  \quad
m(k_{2}) = 2u\cos k_{2}, \quad \mu(k_{2}) = -2t\cos k_{2}.
\end{eqnarray}

\vspace{2mm}

Let us examine the symmetry of our model Lagrangian.
${\cal L}_{0}+{\cal L}^{s}_{I}$ has the $U(1)$ global gauge symmetry under the transformation 
\begin{eqnarray}
\psi_{\alpha}\to e^{i\theta_{V}}\psi_{\alpha},
\end{eqnarray} 
while it has the (1+1)-Lorentz symmetry at $v_{F}k_{F}+\mu(k_{2})=0$.
Of course, the $O(2,1)$-Lorentz symmetry ( including the rotational symmetry in $(x_{1}x_{2})$-plane ) in (2+1)-dimensions is explicitly broken in our model.
From the viewpoint of (1+1)-dimensional Lorentz symmetry, we notice the fact that, 
we should choose the hopping term of the tight-binding picture as a 0th-component of vector, not a scalar, 
for recovering the spectra in Eq. (1) under a special condition of model parameters.
Hence $\mu(k_{2})$ acts as an alternating chemical potential to the system.
This fact becomes more clear when we diagonalize the Lagrangian to obtain the quasiparticle excitation energy spectra, which we will do in Sec. IV.
To compare the effect and/or roles of $\mu(k_{2})$ in the physical property of ${\cal L}$, 
we also added an alternating Dirac mass $m(k_{2})$ as a scalar in (1+1)-dimensions into ${\cal L}_{0}$. 
$m(k_{2})$ is chosen so as to alternate in $k_{2}$-direction.
In terms of $R$ and $L$, ${\cal L}_{0}$ is expressed as
\begin{eqnarray}
{\cal L}_{0} = {\rm tr}\Bigl\{ R^{\dagger}_{\alpha}(i\partial_{0}+iv_{F}\partial_{1}+v_{F}k_{F}+\mu(k_{2}))R_{\alpha} + L^{\dagger}_{\alpha}(i\partial_{0}-iv_{F}\partial_{1}+v_{F}k_{F}+\mu(k_{2}))L_{\alpha} -m(k_{2})[R^{\dagger}_{\alpha}L_{\alpha}+L^{\dagger}_{\alpha}R_{\alpha} ] \Bigr\}.
\end{eqnarray}
Here, ${\rm tr}$ denotes the trace over the spin ( not spinor ) space.
We have observed that, the mass term of $m(k_{2})$ mixes modes of $R_{\alpha}$ and $L_{\alpha}$ ( positive and negative chirality sectors ):
The global $U(1)$ chiral symmetry under the transformation 
\begin{eqnarray}
\psi_{\alpha} \to e^{i\gamma_{5}\theta_{A}}\psi_{\alpha},  \quad
R_{\alpha}\to e^{i\theta_{A}}R_{\alpha}, \quad L_{\alpha} \to e^{-i\theta_{A}}L_{\alpha}, \quad \theta_{A} \in {\bf R},
\end{eqnarray} 
is explicitly broken in ${\cal L}_{0}$ at $m(k_{2})\ne 0$,
while it is symmetric at $m(k_{2})= 0$. 
The hopping term $\mu(k_{2})$ is neutral under the chiral transformation and thus it does not break the symmetry.
The Lagrangian ${\cal L}_{0}+{\cal L}^{s}_{I}$ has the discrete chiral symmetry under 
\begin{eqnarray}
\psi_{\alpha} \to \gamma_{5}\psi_{\alpha}, \quad m(k_{2})\to -m(k_{2}).
\end{eqnarray}
The interaction ${\cal L}^{s}_{I}$ has attractive channels for both particle-hole and particle-particle types at $G_{0}>0$,
and this fact will be shown later by employing the Fierz transformation.
We can consider the origin of ${\cal L}^{s}_{I}$ coming from the particle-phonon coupling as similar to the case of BCS theory~[15]. 
A fermion-fermion repulsive interaction in intrachain 
( with neglecting a possible long-range interaction ) 
similar to the case of the Hubbard interaction can also be treated by the contact interaction at $G_{0}<0$~[28]. 
Throughout this paper, we set aside the question on the mechanisms of the origins of the interactions ${\cal L}^{s}_{I}$, ${\cal L}^{nl-d}_{I}$, and ${\cal L}^{nl-od}_{I}$, 
and examine our model by the attitude of the BCS-NJL ( Nambu, Jona-Lasinio ) theory~[15-26].
The consideration of our method is similar to the important works of Yamaji~[5,6].
Yamaji examined the possibility of coexistence of SDW coming from the Peierls instability and singlet/triplet superconductivity in a quasi-(1+1)-dimensional system,
by employing an unconventional ( = generalized ) BCS interaction into the quasi-(1+1)-dimensional Hubbard model.
( His conclusion is that there is no coexistence of SDW and singlet/triplet superconductivity. )
Our model could be classified as a relativistic version of the Yamaji theory, 
though we do not consider a possible Peierls-type instability of a mode of a finite momentum vector ${\bf Q}$ 
arising from a nesting of the Fermi surface through a response function of our model.
For example, there are many papers to investigate the origin of effective attractive interaction in a particle-particle ( not a particle-hole ) channel for a superconductivity from the Hubbard-type contact repulsive interaction, through the spin fluctuation~[2,8] or the Kohn-Luttinger mechanism~[30]. 
In Ref.~[9-11], the response functions ( namely, the mass terms of order parameters, a collective field is tachyonic or not ) 
were examined for the investigation of the instabilities toward taking place of ordered states.
We have interest more on the physical properties of ordered states than instabilities toward ordered states, 
thus our attitude is closer to the theory of the generalized BCS theory~[17-20] than theories on the mechanism of the origin of an attractive interaction in a superconductivity~[2,8-11].
A possible anisotropy of particle interaction which might be arisen from the inequivalence of the direction $x_{1}$ and $x_{2}$ ( or, $k_{1}$ and $k_{2}$ ) is not taken into account in ${\cal L}^{s}_{I}$ at least at the classical level of our model Lagrangian. 
On the other hand, as mentioned above, we will use ${\cal L}^{nl-d}_{I}$ and ${\cal L}^{nl-od}_{I}$ to contain possible anisotropies of interactions in our model for incorporating the generalized BCS framework in our theory. 
To discuss the symmetries of ${\cal L}^{nl-d}_{I}$ and ${\cal L}^{nl-od}_{I}$, we must specify the structure of potentials $V$ and $W$,
and this will be done in Sec. IV.

\vspace{2mm}

Here, we wish to comment on the relation between our model and several (1+1)-dimensional field theories.
If we omit ${\cal L}_{em}$, ${\cal L}^{nl-d}_{I}$, ${\cal L}^{nl-od}_{I}$ and ${\cal L}_{B}$, 
the (1+1)-dimensional limit of our model will be obtained at $t=u=0$. 
In (1+1)-dimensional field theories, usually the Abelian or the non-Abelian bosonization schema will be performed~[31-37], 
and physical properties of the system are discussed in the context of a realization of the Tomonaga-Luttinger liquid. However, we do not choose this way.
The (1+1)-dimensional relativistic model with including interaction terms of forward scattering is called as the Tomonaga-Luttinger model~[31].
${\cal L}_{0}$ at $t=u=0$ corresponds to the kinetic term of the Tomonaga-Luttinger model.
The Tomonaga-Luttinger model is exactly soluble by the method of bosonization~[31,33], 
and it shows the renormalized gapless ( namely, massless ) linear dispersion $E(k)=v^{*}_{F}|k|$
( $v^{*}_{F}$: A renormalized Fermi velocity ).
The massless Thirring model is also solved by the Abelian bosonization, and it becomes a free Bose theory~[32,37].
In the framework of bosonization, a large class of (1+1)-dimensional models become the quantum sine-Gordon theory. 
In that case, the effects of discrepancies of the models from the Tomonaga-Luttinger model have been examined by renormalization group analysis: 
The results in the renormalization group analysis show that the spectra of several models become gapless ( massless ) when the cosine term is irrelevant, 
while others have gapful ( massive ) spectra when the cosine term of the sine-Gordon theory is relevant. 
${\cal L}_{0}+{\cal L}^{s}_{I}$ at $m(k_{2})=\mu(k_{2})=k_{F}=0$ coincides with the massless Thirring model,
while it becomes a Hubbard-like model at $v_{F}=u=0$.
At $v_{F}\gg t,u$ ( $t,u \gg v_{F}$ ), the kinetic term $\bar{\psi}(-iv_{F}\gamma^{1}\partial_{1})\psi$ 
( the hopping/mass term $\bar{\psi}[-2(u+\gamma^{0}t)\cos k_{2}]\psi$ ) dominates the physical property of the motion of particles of the system. 
It was shown that the massive Thirring model of (1+1)-dimensions becomes a sine-Gordon theory, 
where the mass term of the Thirring model becomes the cosine term of the sine-Gordon theory. 
Physical fermion in the sine-Gordon theory is solition/antisoliton and they become massive under some conditions~[35].  
The massive Thirring model can be converted into the antiferromagnetic $XYZ$-Heisenberg chain through the Jordan-Wigner transformation~[28,29,39].
The two-flavor massless Schwinger model is also mapped into the $S=1/2$ antiferromagnetic spin chain~[42]. 
The massless Schwinger model on the circle $S^{1}$ will have the chiral condensate $\langle \bar{\psi}\psi\rangle\ne 0$~[42].
Moreover, the (1+1)-dimensional massless Gross-Neveu ( Thirring ) model in real space has a $\delta$-function interaction, it was also solved by the Bethe ansatz~[40].
Another important aspect of the (1+1)-dimensional Gross-Neveu model is that it has the asymptotic freedom~[43], 
and it might be an intersting problem for quantum field theoretical point of view that 
how the asymptotic freedom will be modified in our quasi-(1+1)-dimensional model
( the pure-(2+1)-dimensional Gross-Neveu model has no asymptotic freedom ), 
though it may not be important for the low-energy phenomena of our model.
The (1+1)-dimensional Hubbard model describes strongly-correlated electrons where they interact by a short-range ( $\delta$-function ) repulsive potential.
The exact solution of Hubbard model at the half-filling case was obtained by the nested Bethe ansatz, 
and the ground state is a spin-singlet Mott insulator given by a pair condensation of particles and holes~[38]. 
It should be noted that, a particle-hole transformation in the Hubbard interaction gives attractive channels between particles and holes.
Of particular interest in this paper is the possible mechanism of the metal-insulator/metal-superconductor transitions by dynamically generated Dirac mass or Cooper pair condensations.
We use the method of auxiliary fields and the large-$N$ expansion, a kind of Hartree-Fock mean field theory. 
We investigate the effects of the finite interchain hopping and the alternating mass in the dynamical generation of mass-gap and superconductivity.  
Due to the Mermin-Wagner-Coleman theorem, there is no long-range order in (1+1)- and (2+1)- dimensional field theories with short-range interactions~[44-46].
If we evaluate a two-point correlation function, it may show the power low decay, and its infrared behavior is singular~[29]. 
In the theories for examinations of phases/orders in quasi-(1+1) dimensional systems, 
usually the word "the real world is (3+1)-dimensions" has been used for the explanation on this problem~[5,6,9-11]. 
Here, we also obey this attitude.
The fact proved in Ref.~[35] is that, in (1+1)-dimensional four-fermion models, the amplitude of fermion-pair condensation can well develop, 
while the phase ( a bosonic degree of freedom ) of the condensation shows a power-law decay. 
It is just the Kosterlitz-Thouless transition~[47]. 
Thus, at least on the discussion of the development of the amplitude of fermion-pair condensation, 
we believe we can use the large-$N$ mean-field theory.

\vspace{2mm}

We should also mention the relation between our model and other (2+1)-dimensional relativistic theories.
The (2+1)-dimensional Gross-Neveu model has been used for the investigations of dynamical symmetry breaking under various situations,
under finite temperature and density, under the existence of external magnetic field, in Minkowski and curved spacetimes~[25,48-51].
By employing the Nambu$-$Jona-Lasinio model and the Gross-Neveu model,
the relativistic theory of superconductivity in (3+1)-dimensions was extensively studied until now, 
both for examination on the relativistic effects in condensed matters~[22-24] and from the context of color superconductivity in ${\rm QCD}_{4}$~[14].
In Ref.~[25], the relativistic method of superconductivity in (2+1)-dimensions was investigated by using the (2+1)-dimensional Gross-Neveu model, 
for the examination of superconductivity in two-band systems with honeycomb lattice structure, in which case the energy dispersion of electrons/holes is described by
a Dirac Lagrangian.
Our theory considered in this paper can be interpreted as a modification of the (2+1)-dimensional Gross-Neveu model
by introducing the hopping terms and the nonlocal interactions.

\section{The Group Theoretical Consideration of The Order Parameters}

In this section, we consider the symmetry properties of order parameters coming from the condensations of particle-hole and particle-particle pairs.
The main purpose of the group theoretical consideration is to specify suitable order parameters for our discussion of this paper.

First, we examine the symmetry of order parameters in (1+1)-dimensional Lorentz ( spinor ) space.
For this examination, the Fierz transformation of the four-fermion interaction ${\cal L}^{s}_{I}$ is convenient for us,
and the result may also be helpful to handle the nonlocal interactions ${\cal L}^{nl-d}_{I}$ and ${\cal L}^{nl-od}_{I}$.
By utilizing the complete set $\{1,\gamma^{\mu},\gamma_{5}\}$ ( $\mu=0,1$ ) in $2\times 2$ matrix spinor space, 
the Fierz rearrangement in the four-fermion interaction was obtained into the following form:
\begin{eqnarray}
(\bar{\psi}\psi)^{2} &=& \frac{1}{2}\Bigl[
-(\bar{\psi}\psi)(\bar{\psi}\psi)
-(\bar{\psi}\gamma^{0}\psi)(\bar{\psi}\gamma^{0}\psi)
+(\bar{\psi}\gamma^{1}\psi)(\bar{\psi}\gamma^{1}\psi)
-(\bar{\psi}\gamma_{5}\psi)(\bar{\psi}\gamma_{5}\psi)
\Bigr]   \nonumber \\
&=& \frac{1}{2}\Bigl[ 
  (\bar{\psi}\gamma_{5}C\bar{\psi}^{T})(\psi^{T}C^{-1}\gamma_{5}\psi) 
+ (\bar{\psi}\gamma^{0}\gamma_{5}C\bar{\psi}^{T})(\psi^{T}C^{-1}\gamma_{5}\gamma^{0}\psi)    \nonumber \\
& & - (\bar{\psi}\gamma^{1}\gamma_{5}C\bar{\psi}^{T})(\psi^{T}C^{-1}\gamma_{5}\gamma^{1}\psi) 
+ (\bar{\psi}C\bar{\psi}^{T})(\psi^{T}C^{-1}\psi) \Bigr].
\end{eqnarray}
Here, the spin index $\alpha$ of $\psi$-field is not written explicitly for simplicity.
Unlike the case of (3+1)-dimensions, here we do not have any axial-vector channels like $\bar{\psi}\gamma^{\mu}\gamma_{5}\psi$,
because the matrix space is four-dimensions and the linear independent set for the expansion of the space will be constructed only by scalar, vector and pseudoscalar.
By decomposing the Dirac field $\psi$ into the left and right moving modes under our convention of the gamma matrices, one finds the following expressions
for densities of scalar $\varrho_{S}$, vector $\varrho^{\mu}_{V}$ ( $\mu=0,1$ ) and pseudoscalar $\varrho_{P}$ ( "pion" ) of fermion-antifermion pair types 
( "diagonal" part of the density matrix, CDW/SDW types ):
\begin{eqnarray}
\varrho_{S}(x) &\equiv& \bar{\psi}(x)\psi(x) = L^{\dagger}(x)R(x) + R^{\dagger}(x)L(x),  \\
\varrho^{0}_{V}(x) &\equiv& \bar{\psi}(x)\gamma^{0}\psi(x) = R^{\dagger}(x)R(x) + L^{\dagger}(x)L(x),  \\
\varrho^{1}_{V}(x) &\equiv& \bar{\psi}(x)\gamma^{1}\psi(x) = R^{\dagger}(x)R(x) - L^{\dagger}(x)L(x),  \\
\varrho_{P}(x) &\equiv& \bar{\psi}(x)i\gamma_{5}\psi(x) = iL^{\dagger}(x)R(x) - iR^{\dagger}(x)L(x).
\end{eqnarray}
The non-vanishing VEV ( vacuum expectation value ) of $\varrho_{S}$ ( chiral condensation ) indicates the realization of spontaneous chiral symmetry breaking of the model ${\cal L}_{0}+{\cal L}^{s}_{I}$,
while a non-zero VEV of $\varrho_{P}$ can be called as a kind of "pion" condensation.
It is clear from these expressions, the interaction $(\bar{\psi}\psi)^{2}$ includes the Umklapp processes $R^{\dagger}R^{\dagger}LL$ and $L^{\dagger}L^{\dagger}RR$,
while they will be removed by taking the combination $(\bar{\psi}\psi)^{2}+(\bar{\psi}i\gamma_{5}\psi)^{2}$:
The global $U(1)$ chiral symmetric interaction does not have the Umklapp processes.
On the other hand, fermion-fermion pair densities of scalar $\xi_{S}$, vector $\xi^{\mu}_{V}$ ( $\mu=0,1$ ) and pseudoscalar $\xi_{P}$
( "off diagonal" part of the density matrix, Cooper pair types ) become
\begin{eqnarray}
\xi_{S}(x) &\equiv& \psi^{T}(x)C^{-1}\gamma_{5}\psi(x) = -L(x)R(x)-R(x)L(x), \\
\xi^{0}_{V}(x) &\equiv& \psi^{T}(x)C^{-1}\gamma_{5}\gamma^{0}\psi(x) = -R(x)R(x)-L(x)L(x), \\
\xi^{1}_{V}(x) &\equiv& \psi^{T}(x)C^{-1}\gamma_{5}\gamma^{1}\psi(x) = -R(x)R(x)+L(x)L(x), \\
\xi_{P}(x) &\equiv& \psi^{T}(x)C^{-1}i\psi(x) = -iL(x)R(x)+iR(x)L(x),
\end{eqnarray}
and
\begin{eqnarray}
\overline{\xi}_{S}(x) &\equiv& \bar{\psi}(x)\gamma_{5}C\bar{\psi}^{T}(x) = -R^{\dagger}(x)L^{\dagger}(x)-L^{\dagger}(x)R^{\dagger}(x) = (\xi_{S})^{\dagger}, \\
\overline{\xi}^{0}_{V}(x) &\equiv& \bar{\psi}(x)\gamma^{0}\gamma_{5}C\bar{\psi}^{T}(x) = -L^{\dagger}(x)L^{\dagger}(x) - R^{\dagger}(x)R^{\dagger}(x) = (\xi^{0}_{V})^{\dagger}, \\
\overline{\xi}^{1}_{V}(x) &\equiv& \bar{\psi}(x)\gamma^{1}\gamma_{5}C\bar{\psi}^{T}(x) = L^{\dagger}(x)L^{\dagger}(x)-R^{\dagger}(x)R^{\dagger}(x) = (\xi^{1}_{V})^{\dagger}, \\
\overline{\xi}_{P}(x) &\equiv& \bar{\psi}(x)iC\bar{\psi}^{T}(x) = iR^{\dagger}(x)L^{\dagger}(x)-iL^{\dagger}(x)R^{\dagger}(x) = (\xi_{P})^{\dagger}.
\end{eqnarray}
The reason of the identification of the symmetries of scalar $S$, vector $V$ and pseudoscalar $P$ of the Cooper pairs will be clarified later in this section.
If the fields $R$ and $L$ are spinless, the vectorial Cooper pairs $\xi^{0}_{V}$ and $\xi^{1}_{V}$ are forbidden by the Pauli principle.
$\varrho_{S}$ and $\varrho_{P}$ break the $U(1)$ chiral symmetry and keep the $U(1)$ gauge symmetry, 
while $\xi_{S}$ and $\xi_{P}$ are chiral invariant and break the $U(1)$ gauge symmetry. 
${\cal L}^{s}_{I}$ is given in a quadratic form of these densities in general.

\vspace{2mm}

To understand the physical characters of ordered states, 
we examine the transformation properties under the charge-conjugation ${\cal C}$, spatial inversion ${\cal P}$ and time reversal ${\cal T}$ symmetries.
Especially, the broken ${\cal T}$ is interesting with respect to the relation with spontaneous magnetization of the system.
The parity transformation in (1+1)-dimensions is given by
\begin{eqnarray}
\psi(x_{0},x_{1}) \stackrel{\cal P}{\to} \gamma^{0}\psi(x_{0},-x_{1}), \quad \bar{\psi}(x_{0},x_{1}) \stackrel{\cal P}{\to} \bar{\psi}(x_{0},-x_{1})\gamma^{0}.
\end{eqnarray}
Therefore one finds
\begin{eqnarray}
& & \varrho_{S}(x_{0},x_{1}) \stackrel{\cal P}{\to} \varrho_{S}(x_{0},-x_{1}),  \quad 
\varrho^{0}_{V}(x_{0},x_{1}) \stackrel{\cal P}{\to} \varrho^{0}_{V}(x_{0},-x_{1}),  \nonumber \\
& & \varrho^{1}_{V}(x_{0},x_{1}) \stackrel{\cal P}{\to} -\varrho^{1}_{V}(x_{0},-x_{1}), \quad 
\varrho_{P}(x_{0},x_{1}) \stackrel{\cal P}{\to} -\varrho_{P}(x_{0},-x_{1}),   \\
& & \xi_{S}(x_{0},x_{1}) \stackrel{\cal P}{\to} \xi_{S}(x_{0},-x_{1}),  \quad 
\xi^{0}_{V}(x_{0},x_{1}) \stackrel{\cal P}{\to} \xi^{0}_{V}(x_{0},-x_{1}),  \nonumber \\
& & \xi^{1}_{V}(x_{0},x_{1}) \stackrel{\cal P}{\to} -\xi^{1}_{V}(x_{0},-x_{1}), \quad 
\xi_{P}(x_{0},x_{1}) \stackrel{\cal P}{\to} -\xi_{P}(x_{0},-x_{1}).
\end{eqnarray}
Hence, $\varrho_{S}$ and $\xi_{S}$ are parity-even.
Under the charge conjugation,
\begin{eqnarray}
& & \varrho_{S} \stackrel{\cal C}{\to} \varrho_{S},  \quad 
\varrho^{0}_{V} \stackrel{\cal C}{\to} -\varrho^{0}_{V},  \quad
\varrho^{1}_{V} \stackrel{\cal C}{\to} -\varrho^{1}_{V}, \quad 
\varrho_{P} \stackrel{\cal C}{\to} -\varrho_{P},   \\
& & \xi_{S} \stackrel{\cal C}{\to} -\bar{\xi}_{S},  \quad 
\xi^{0}_{V} \stackrel{\cal C}{\to} \bar{\xi}^{0}_{V},  \quad
\xi^{1}_{V} \stackrel{\cal C}{\to} \bar{\xi}^{1}_{V}, \quad 
\xi_{P} \stackrel{\cal C}{\to} -\bar{\xi}_{P}.
\end{eqnarray}
The time-reversal transformation ${\cal T}$ is defined by
\begin{eqnarray}
\psi(x_{0},x_{1}) \stackrel{\cal T}{\to} \gamma^{0}\psi(-x_{0},x_{1}), \quad \bar{\psi}(x_{0},x_{1}) \stackrel{\cal T}{\to} \bar{\psi}(-x_{0},x_{1})\gamma^{0},
\end{eqnarray}
with the rule to take the complex conjugation to matrix elements ( anti-unitary operations ).
The densities will be transformed as
\begin{eqnarray}
& & \varrho_{S}(x_{0},x_{1}) \stackrel{\cal T}{\to} \varrho^{*}_{S}(-x_{0},x_{1}),  \quad 
\varrho^{0}_{V}(x_{0},x_{1}) \stackrel{\cal T}{\to} \varrho^{0*}_{V}(-x_{0},x_{1}),  \nonumber \\
& & \varrho^{1}_{V}(x_{0},x_{1}) \stackrel{\cal T}{\to} -\varrho^{1*}_{V}(-x_{0},x_{1}), \quad 
\varrho_{P}(x_{0},x_{1}) \stackrel{\cal T}{\to} -\varrho^{*}_{P}(-x_{0},x_{1}),   \\
& & \xi_{S}(x_{0},x_{1}) \stackrel{\cal T}{\to} \xi^{*}_{S}(-x_{0},x_{1}),  \quad 
\xi^{0}_{V}(x_{0},x_{1}) \stackrel{\cal T}{\to} \xi^{0*}_{V}(-x_{0},x_{1}),  \nonumber \\
& & \xi^{1}_{V}(x_{0},x_{1}) \stackrel{\cal T}{\to} -\xi^{1*}_{V}(-x_{0},x_{1}), \quad 
\xi_{P}(x_{0},x_{1}) \stackrel{\cal T}{\to} -\xi^{*}_{P}(-x_{0},x_{1}).
\end{eqnarray}

\vspace{2mm}

In considerations on superconducting order parameters, one has to find both its transformation property under several groups and the Pauli principle~[20].
In the case of this paper, especially we should consider the spin and the (1+1)-dimensional Lorentz ( namely, spinor ) symmetries.
( We must not confuse the "spin" and "spinor" degrees of freedom. The spin degree is introduced as a kind of isospin in our model, while the spinor degree has the origin in (1+1)-Lorentz symmetry. )
The superconducting order parameter will be defined as follows:
\begin{eqnarray}
\Delta(x) \sim \psi(x)\psi^{T}(x), \quad \overline{\Delta}(x) \sim \bar{\psi}^{T}(x)\bar{\psi}(x), \quad 
\overline{\Delta} = \gamma^{0}\Delta^{\dagger}\gamma^{0}, \quad \Delta \stackrel{\cal C}{\to} C\overline{\Delta}C^{-1}, 
\quad \overline{\Delta} \stackrel{\cal C}{\to} C\Delta C^{-1}.
\end{eqnarray}
Here, $\Delta$ and $\overline{\Delta}$ are proportional to $G_{0}$, and we omit it for the notational simplicity of the discussion in this section.
$\Delta$ and $\overline{\Delta}$ have $4\times 4$ matrix structures coming from the direct product of the two-dimensional spin and the two-dimensional spinor spaces. 
The (1+1)-dimensional Lorentz transformation is determined by the consideration of Lorentz algebra in the following form:
\begin{eqnarray}
\psi &\to& S\psi, \quad S\equiv \exp\Bigl( -\frac{i}{4}\epsilon_{\mu\nu}\sigma^{\mu\nu} \Bigr), \quad
\epsilon_{\mu\nu} = -\epsilon_{\nu\mu}, \quad \sigma^{\mu\nu} = \frac{i}{2}[\gamma^{\mu},\gamma^{\nu}],    \nonumber \\
S &=& \exp\Bigl(\frac{\omega}{2} \gamma_{5} \Bigr), \quad \omega = \epsilon_{01}, \quad \omega\in{\bf R}, \quad S^{-1} = \gamma^{0}S^{\dagger}\gamma^{0}.
\end{eqnarray}
The definition of $S^{-1}$ is crucial to keep the quantity like $\bar{\psi}\psi$ as a scalar.
Under the (1+1)-dimensional Lorentz transformation, the superconducting order parameter will be transformed as
\begin{eqnarray}
\Delta &\sim& \psi\psi^{T} \to S\psi \psi^{T}S^{T} = e^{\frac{\omega}{2}\gamma_{5}}\psi\psi^{T}e^{\frac{\omega}{2}\gamma^{T}_{5}} = \psi\psi^{T} + \frac{\omega}{2}\Bigl( \psi\psi^{T}\gamma_{5} + \gamma_{5}\psi\psi^{T}\Bigr) + {\cal O}(\omega^{2}).
\end{eqnarray}
Hence, expanding $\Delta$ by the gamma matrices in (1+1)-dimensions as a complete set in $2\times 2$ matrix spinor space,
one finds
\begin{eqnarray}
\Delta = \Bigl(\Delta_{S}+\Delta^{\mu}_{V}\gamma^{\mu}+\Delta_{P}i\gamma_{5}\Bigr)C^{-1}\gamma_{5}, \quad 
\overline{\Delta} = \gamma_{5}C\Bigl(\overline{\Delta}_{S}+\overline{\Delta}^{\mu}_{V}\gamma^{\mu}+\overline{\Delta}_{P}(-i\gamma_{5})\Bigr).
\end{eqnarray}
Here, $S$ ,$V$ and $P$ denote scalar, vector, and pseudoscalar, respectively.
The scalar $\Delta_{S}C\gamma_{5}$ has been chosen so as to be a (1+1)-dimensional parity-even order parameter under (32). 
( The correspondences between the components of $\Delta$ and $\xi_{S}$, $\xi^{\mu}_{V}$, $\xi_{P}$ are obvious for us:
$\xi_{S}\leftrightarrow\Delta_{S}C^{-1}\gamma_{5}$, so forth. )
On the other hand, under the rotation of $SU(2)$ spin space, the Dirac field is transformed as ( $\tau_{i}$ ( $i=1,2,3$ ); the Pauli matrices )
\begin{eqnarray}
\psi &\to& \exp\Bigl( \frac{i}{2}\theta_{i}\tau_{i} \Bigr)\psi, \quad \theta_{i}\in{\bf R}, \quad \tau^{T}_{i} = -\tau_{2}\tau_{i}\tau_{2}.
\end{eqnarray}
The order parameter matrix is transformed as 
\begin{eqnarray}
\Delta \sim \psi\psi^{T} \to e^{\frac{i}{2}\theta_{i}\tau_{i}}\psi\psi^{T}e^{\frac{i}{2}\theta_{i}\tau^{T}_{i}} = \psi\psi^{T} + \frac{i}{2}\theta_{i}[\tau_{i},\psi\psi^{T}\tau_{2}]\tau_{2} + {\cal O}(\theta^{2}_{i}).
\end{eqnarray}
Therefore, we find the following expansion appropriate to define spin singlet and triplet Cooper pairs:
\begin{eqnarray}
\Delta &=& \Bigl( \Delta^{(1)} + \Delta^{(3)}_{i}\tau_{i} \Bigr)i\tau_{2}, \quad 
\overline{\Delta} = -i\tau_{2}\Bigl( \overline{\Delta^{(1)}} + \overline{\Delta^{(3)}_{i}}\tau_{i} \Bigr), \quad ( i = 1,2,3 ).
\end{eqnarray}
Here, $\Delta^{(1)}$ and $\Delta^{(3)}_{i}$ denote order parameters of spin singlet and triplet Cooper pairs, respectively. 
The spin singlet pairing of any Lorentz symmetry ( $S$, $V$, or $P$ ) becomes a scalar while the triplet has the vectorial nature under the spin rotation:
Namely, the spin ( or, flavor ) and spinor ( Lorentz ) degrees of freedom are completely decoupled in our model.
Due to the Pauli principle, the matrix $\Delta$ must change the sign under the transposition in spinor$\otimes$spin space:
\begin{eqnarray}
\Delta^{T} = -\Delta, \quad \overline{\Delta}^{T} = - \overline{\Delta}.
\end{eqnarray}
By taking into account both the spin and Lorentz ( the (1+1)-dimensional spinor space ) degrees of freedom, we find
the spin-singlet scalar pairing $\Delta^{(1)}_{S}(x)C\gamma_{5}\otimes i\tau_{2}$ is allowed by the Pauli principle,
and it corresponds to the s-wave BCS state. 
In this case $\Delta^{(1)}_{S}(x)=\Delta^{(1)}_{S}(-x)$ ( an even function under parity ) has to be satisfied. 
On the contrary, the spin-triplet scalar ( a vector in spin space ) $\overrightarrow{\Delta^{(3)}_{S}}(x)\cdot C\gamma_{5}\otimes i\vec{\tau}\tau_{2}$ has to become an odd-parity function under the spacial inversion to satisfy the Pauli principle: $\overrightarrow{\Delta^{(3)}_{S}}(x)=-\overrightarrow{\Delta^{(3)}_{S}}(-x)$. 
If our model does not have the spin degree of freedom, only the pseudoscalar ( not the scalar ) component of the Cooper pair is allowed by the Pauli principle.
A linear combination of $\psi^{T}C^{-1}\gamma_{5}\gamma^{0}\otimes i\tau_{2}\psi$ and $\bar{\psi}\gamma^{0}\gamma_{5}C\otimes i\tau_{2}\bar{\psi}^{T}$ 
corresponds to the Majorana-like mass term as $Ri\tau_{2}R+R^{\dagger}i\tau_{2}R^{\dagger}$ and $Li\tau_{2}L+L^{\dagger}i\tau_{2}L^{\dagger}$, and it cannot exist if our model is spinless.
In (3+1)-dimensional field theory, a superconducting order parameter is a specific linear combination of ( parity breaking ) Majorana mass terms~[23,24].
Usually, the symmetry of Cooper pairs will be defined in the momentum space, while $\Delta(x)$ of a homogeneous system cannot have a momentum-dependence
and hence it cannot become an odd-parity function in momentum space, thus spin-triplet Cooper pairs must be handled by the generalized BCS framework.

\vspace{2mm}

Based on the discussion given above, we prepare the following interaction for generating both the $s$-wave dynamical Dirac mass ( coming from a nonzero VEV of $\varrho_{S}$ ) and spin-singlet $s$-wave superconductivity ( again, from a nonzero VEV of $\xi_{S}$ ) from ${\cal L}^{s}_{I}$ in our theory:
\begin{eqnarray}  
{\cal L}^{s}_{I} &=& \frac{G_{0}}{2N}(\bar{\psi}\psi)^{2} = \frac{G_{0}}{4N}(\bar{\psi}\psi)^{2} + \frac{G_{0}}{8N}(\bar{\psi}i\tau_{2}\gamma_{5}C\bar{\psi}^{T})(\psi^{T}C^{-1}\gamma_{5}i\tau_{2}\psi).
\end{eqnarray}
Here, we only keep the relevant terms for the dynamical generation of Dirac mass in a particle-hole channel and scalar ( in (1+1)-dimensional Lorentz space ) Cooper pairs in a particle-particle channel.
The other terms appear in the Fierz transformation (??) may act as quantum fluctuations around a stationary point in the ( path integral ) quantization of our theory,
and they are suppressed by the large-$N$ limit because propagators of the fluctuating collective modes are proportional to $1/N$ inside the effective action of the theory.
This interaction can only handle $s$-wave pairings. For the generalization of $p$-, $d$- and $f$- wave pairings, especially to consider spin-triplet Cooper pairs, 
we will give the generalized BCS framework in Sec. IV.

\vspace{2mm}

Another important aspect of our model under the context of the purpose of this paper is the Pauli-G\"{u}rsey ( PG ) symmetry, 
the symmetry of a rotation in particle-antiparticle space~[51-53].
The transformation of the PG symmetry in (1+1)-dimensions is determined as follows~[51]:
\begin{eqnarray}
\psi &\to& P_{L}\psi + P_{R}\psi^{c} = \left(
\begin{array}{c}
-R^{*} \\
L
\end{array}
\right), \quad \bar{\psi} \to \bar{\psi}P_{R} + \overline{\psi^{c}}P_{L} = (L^{\dagger},R),   \nonumber \\
\psi^{c} &\to& P_{L}\psi^{c} + P_{R}\psi = \left(
\begin{array}{c}
R \\
L^{*}
\end{array}
\right), \quad \overline{\psi^{c}} \to \overline{\psi^{c}}P_{R} + \bar{\psi}P_{L} = (-L,R^{\dagger}),
\end{eqnarray}
where,
\begin{eqnarray}
P_{L} &\equiv& \frac{1}{2}(1-\gamma_{5}), \quad P_{R} \equiv \frac{1}{2}(1+\gamma_{5}), \quad
P_{L}+P_{R} = 1, \quad P_{L}P_{L} = P_{L}, \quad P_{R}P_{R} = P_{R}, \quad P_{L}P_{R} = P_{R}P_{L} = 0,  \nonumber \\
\psi^{c} &=& C\bar{\psi}^{T} = \left(
\begin{array}{c}
-R^{*} \\
L^{*}
\end{array}
\right), \quad \overline{\psi^{c}} = -\psi^{T}C^{-1} = (-L,R).
\end{eqnarray}
Here, the definition of the charge-conjugation of $\psi$ has been given in terms of $R$ and $L$ for the convenience of our discussion. 
( More compact expression will be given in the form of displacement operators:
\begin{eqnarray}
Q &\equiv& \left(
\begin{array}{c}
\psi \\
\psi^{c} 
\end{array}
\right), \quad \widetilde{Q} \equiv (\overline{\psi^{c}},\bar{\psi}), \quad Q' = \exp[i\varepsilon\widehat{F}]Q, \quad \widetilde{Q'} = \widetilde{Q}\exp[-i\varepsilon\widehat{F}], \quad
\widehat{F} \equiv \left(
\begin{array}{cc}
P_{L} & P_{R} \\
P_{R} & P_{L}
\end{array}
\right),\quad \widetilde{Q'}Q' = \widetilde{Q}Q.
\end{eqnarray}
The quantity $\widetilde{Q}Q$, can be called as a "norm" or a "inner product", is "conserved" under the PG transformation. )
Hence, in our definition of $\psi$ in this paper, the PG transformation corresponds to the additions of the complex conjugates $R^{*}$ or $L^{*}$ 
to a specific chirality sector $R$ or $L$ of $\psi$.
In order to examine the condition of the realization of the PG symmetry in our model, we rewrite the kinetic term ${\cal L}_{0}$ in the following form:
\begin{eqnarray}
{\cal L}_{0} &=& \bar{\psi}_{\alpha}\Bigl(\gamma^{0}\{ i\partial_{0} + v_{F}k_{F}+\mu(k_{2}) \}+iv_{F}\gamma^{1}\partial_{1} -m(k_{2})  \Bigr)\psi_{\alpha}   \nonumber \\
&=& \overline{\psi^{c}}_{\alpha}
\Bigl(\gamma^{0}\{ i\partial_{0} - v_{F}k_{F}-\mu(k_{2}) \}+iv_{F}\gamma^{1}\partial_{1} -m(k_{2})  \Bigr)\psi^{c}_{\alpha} \equiv {\cal L}^{c}_{0}.
\end{eqnarray}
By performing the PG transformation defined in (49) to ${\cal L}_{0}=({\cal L}_{0}+{\cal L}^{c}_{0})/2$, we find that
the Lagrangian ${\cal L}_{0}$ has the PG symmetry at $v_{F}k_{F}+\mu(k_{2})=0$, which is the same with the Lorentz symmetry condition for ${\cal L}_{0}$, 
while the mass term $m(k_{2})$ is neutral under the PG transformation.
The points as solutions of the equation $0=v_{F}k_{F}+\mu(k_{2})$ in the $(k_{1},k_{2})$-surface of the energy eigenvalue may exist under the condition $|v_{F}k_{F}/(2t)|\le 1$:
At $k_{2}=\cos^{-1}(v_{F}k_{F}/(2t))$, we cannot distinguish between $\psi$ and $\psi^{c}$, $\bar{\psi}$ and $\overline{\psi^{c}}$:
In this case, for example, $R^{\dagger}L$ and $RL$ are the same composite field.
We should note that this PG-symmetric condition has been obtained in the case of "free" field theory, 
and interactions ( namely, renormalization ) between particles would modify it in general. 
The densities will be transformed under the PG group as follows:
\begin{eqnarray}
& & \delta(\bar{\psi}\psi) = \bar{\psi}P_{R}\psi^{c} + \overline{\psi^{c}}P_{L}\psi,   \quad
\delta(\bar{\psi}\gamma^{0}\psi) =  \bar{\psi}\gamma^{0}\psi,   \quad
\delta(\bar{\psi}\gamma^{1}\psi) =  \bar{\psi}\gamma^{1}\psi,   \quad
\delta(\bar{\psi}\gamma_{5}\psi) =  \bar{\psi}P_{R}\psi^{c} - \overline{\psi^{c}}P_{L}\psi,    \nonumber \\
& & \delta(\psi^{T}C^{-1}\gamma_{5}\psi) = 0,   \quad
\delta(\psi^{T}C^{-1}\gamma_{5}\gamma^{0}\psi) = -\psi^{T}P_{L}\psi -\bar{\psi}P_{L}\bar{\psi}^{T},   \nonumber \\
& & \delta(\psi^{T}C^{-1}\gamma_{5}\gamma^{1}\psi) = \psi^{T}P_{L}\psi -\bar{\psi}P_{L}\bar{\psi}^{T},   \quad
\delta(\psi^{T}C^{-1}\psi) = -2\bar{\psi}P_{L}\psi,   \nonumber \\
& & \delta(\bar{\psi}\gamma_{5}C\bar{\psi}^{T}) = 0,   \quad
\delta(\bar{\psi}\gamma^{0}\gamma_{5}C\bar{\psi}^{T}) = -\psi^{T}P_{R}\psi -\bar{\psi}P_{R}\bar{\psi}^{T},   \nonumber \\
& & \delta(\bar{\psi}\gamma^{1}\gamma_{5}C\bar{\psi}^{T}) = -\psi^{T}P_{R}\psi + \bar{\psi}P_{R}\bar{\psi}^{T},   \quad
\delta(\bar{\psi}C\bar{\psi}^{T}) = 2\bar{\psi}P_{R}\psi.
\end{eqnarray}
From these results, we find that the four-fermion interaction $(\psi^{T}C^{-1}\gamma_{5}\psi)(\bar{\psi}\gamma_{5}C\bar{\psi}^{T})$ is explicitly PG invariant
( $\psi^{T}C^{-1}\gamma_{5}\psi$ and $\bar{\psi}\gamma_{5}C\bar{\psi}^{T}$ are PG singlet ).
The displacements of the scalar $\delta(\bar{\psi}\psi)$ and the pseudoscalar $\delta(\bar{\psi}\gamma_{5}\psi)$ show Cooper-pair-type fermion-fermion pairs.
On the contrary, the vectorial fermion-antifermion ( fermion-fermion ) pairs give again fermion-antifermion ( fermion-fermion ) pairs under the PG transformation.

\vspace{2mm}

Before closing this section, we should also mention the Vafa-Witten theorem~[55]. 
This theorem states that the isospin ( flavor ) symmetry cannot be broken in Nambu$-$Jona-Lasinio-type four-fermion-interaction models through the chiral condensation.
However, this theorem was proved by using a set of Lorentz invariant eigenvalues by solving the two-flavor Dirac equations
$i\Dfey q=\lambda q$ ( $q=u,d$ quarks ).
Because the Lorentz symmetry is explicitly broken in our model, the theorem could not be applied to our theory naively.
For example, we consider the possibility for spontaneous generation of the vectorial condensate $\bar{\psi}_{\alpha}(\sigma_{i})_{\alpha\beta}\psi_{\beta}$ ( a kind of SDW ) which could break the rotational symmetry in the spin space of our model. 
Furthermore, the existence of the external magnetic field $\vec{B}$ will explicitly break the spin symmetry in our theory.

\section{The Effective Potential and The Gap Equations}

In this section, we will obtain the effective potential and the gap equations of several order parameters.
By taking into account the results in the group theoretical consideration on the order parameters given in the previous section,
we only consider the scalar components of the (1+1)-Lorentz symmetry.
By using local ( $\sigma(x)$, $\Delta^{(1)}_{S}(x)$ ) and bilocal ( $\chi_{S}(x,y)$, $\Xi_{S}(x,y)$ ) auxiliary fields~[56,57], the Lagrangian becomes
\begin{eqnarray}
{\cal L} &=& \Bigl( -\frac{N}{G_{0}}\sigma^{2}(x) -\frac{2N}{G_{0}}|\Delta^{(1)}_{S}(x)|^{2} \Bigr)\delta_{xy} \nonumber \\
& & -\frac{N}{2}\chi_{S}(x,y) V^{-1}(x,y;x',y')\chi_{S}(x',y') -\frac{N}{2}\overline{\Xi_{S}}(x,y) W^{-1}(x,y;x',y')\Xi_{S}(x',y') + \frac{1}{2}\bar{\Psi}(x){\cal M}(x,y)\Psi(y).
\end{eqnarray}
Here, $\Delta^{(1)}_{S}$ is the scalar, $s$-wave (iso)spin-singlet component of the order parameter $\Delta$.
$\chi_{S}(x,y)$ and $\Xi_{S}(x,y)$ are scalar, proportional to the bilinear forms $\bar{\psi}(x)(1+\vec{\tau})\psi(y)$ and $\psi^{T}(x)C^{-1}\gamma_{5}(1+\vec{\tau})i\tau_{2}\psi(y)$,
respectively. They include both spin-singlet and triplet pairs in general.
$\sigma$ and $\chi_{S}$ have no $U(1)$-phase degree of freedom.
In the expression of the Lagrangian, we have introduced the eight-component Nambu notation~[16,22-25] in the following definition: 
\begin{eqnarray}
\Psi(x) \equiv \left(
\begin{array}{c}
\psi_{\alpha}(x) \\
\bar{\psi}^{T}_{\alpha}(x)
\end{array}
\right), \quad \bar{\Psi}(x) = (\bar{\psi}_{\alpha}(x),\psi^{T}_{\alpha}(x)),
\end{eqnarray}
and the matrix ${\cal M}$ is defined as follows:
\begin{eqnarray}
& & {\cal M} = {\cal M}^{(1)}(x)\delta_{xy} + {\cal M}^{(2)}(x,y), \quad 
{\cal M}^{(1)} =
\left(
\begin{array}{cc}
{\cal M}_{11} & {\cal M}_{12} \\
{\cal M}_{21} & {\cal M}_{22}
\end{array}
\right),  \quad 
{\cal M}^{(2)}(x,y) \equiv \left(
\begin{array}{cc}
-\chi_{S}(x,y) & \Xi_{S}(x,y)   \\
\overline{\Xi_{S}}(x,y) & \chi_{S}(x,y)
\end{array}
\right), 
\end{eqnarray}
where,
\begin{eqnarray}
& & {\cal M}_{11} \equiv \gamma^{0}\{ i\partial_{0}+v_{F}k_{F}+\mu(k_{2})-\frac{g_{J}\mu_{B}}{2}B_{3}\tau_{3}\} +iv_{F}\gamma^{1}\partial_{1} -\sigma -m(k_{2}), \\
& & {\cal M}_{12} \equiv \Delta^{(1)}_{S}(x) C^{-1}\gamma_{5}\otimes i\tau_{2}, \quad {\cal M}_{21} \equiv -\overline{\Delta^{(1)}_{S}}(x)\gamma_{5}C\otimes i\tau_{2}, \\
& & {\cal M}_{22} \equiv \gamma^{0T}\{ i\partial_{0}-v_{F}k_{F}-\mu(k_{2})+\frac{g_{J}\mu_{B}}{2}B_{3}\tau_{3}\} +iv_{F}\gamma^{1T}\partial_{1} +\sigma + m(k_{2}),   \\
& & \chi_{S}(x,y) \equiv \chi^{(1)}_{S}(x,y) + \overrightarrow{\chi^{(3)}_{S}}(x,y)\cdot\vec{\tau}, \\
& & \Xi_{S}(x,y) \equiv \Bigl( \Xi^{(1)}_{S}(x,y) + \overrightarrow{\Xi^{(3)}_{S}}(x,y)\cdot\vec{\tau} \Bigr)C^{-1}\gamma_{5}\otimes i\tau_{2}, \quad \overline{\Xi_{S}}(x,y) \equiv -i\tau_{2}\otimes\gamma_{5}C\Bigl( \overline{\Xi^{(1)}_{S}}(x,y)+\overrightarrow{\overline{\Xi^{(3)}_{S}}}(x,y)\cdot\vec{\tau} \Bigr).
\end{eqnarray}
Here, we have presented ${\cal M}$ in the form of Fourier-transformed expression in $x_{2}$-coordinate.
$\chi^{(1)}_{S}$ and $\Xi^{(1)}_{S}$ denote the spin singlet pairs, while $\overrightarrow{\chi^{(3)}_{S}}$ and $\overrightarrow{\Xi^{(3)}_{S}}$ are triplet pairs.
The Fourier transforms of bilocal auxiliary fields and the nonlocal interactions become 
\begin{eqnarray}
& & \chi_{S}(x,y) = \int_{P}\int_{q}e^{-iPX-iqr}\chi_{S}(P,q), \quad
\Xi_{S}(x,y) = \int_{P}\int_{q}e^{-iPX-iqr}\Xi_{S}(P,q), \quad
\overline{\Xi_{S}}(x,y) = \int_{P}\int_{q}e^{iPX+iqr}\overline{\Xi_{S}}(P,q),   \nonumber \\
& & V(x,y;x',y') = \int_{P}\int_{P'}\int_{q}\int_{q'} e^{-iPX-iP'X'-iqr-iq'r'}V(P,q;P',q'),  \nonumber \\
& & W(x,y;x',y') = \int_{P}\int_{P'}\int_{q}\int_{q'} e^{-iPX-iP'X'-iqr-iq'r'}W(P,q;P',q'),  \nonumber \\
& & X \equiv \frac{x+y}{2}, \quad X' \equiv \frac{x'+y'}{2}, \quad r \equiv x-y, \quad r' \equiv x'-y'.
\end{eqnarray}
Here, we have used the abbreviation defined in Eq. (13).
$X=(X_{0},X_{1},X_{2})$ denote the center of mass coordinates for fermion-antifermion and fermion-fermion pairs, 
while $r=(x_{0}-y_{0},x_{1}-y_{1},x_{2}-y_{2})$ are the relative coordinates between particles they would form a pair condensation.
Hereafter, we take $P_{\mu}=P'_{\mu}=0$, choose the total momentum of any Cooper pairs at the origin of $(k_{1},k_{2})$-space.
The partition function $Z$ considered in this paper is defined as follows:
\begin{eqnarray}
Z &=& \int {\cal D}\psi{\cal D}\bar{\psi}{\cal D}\sigma{\cal D}\Delta^{(1)}_{S}{\cal D}\overline{\Delta^{(1)}_{S}}{\cal D}\chi_{S}{\cal D}\Xi_{S}{\cal D}\overline{\Xi_{S}}\exp\Bigl[ i\sum_{n_{x}}\int dx_{0}dx_{1} \widetilde{\cal L}\Bigr]. 
\end{eqnarray}
It seems needless to say, however, we wish to emphasize that, this functional integral contains all of the information on the dynamics of the system.
We employ the steepest descent approximation to the integration of the auxiliary fields in $Z$. 
The effective action $\Gamma$ is found to be
\begin{eqnarray}
\Gamma &=& -\sum_{n_{x}}\int dx_{0}dx_{1} \Bigg[ \Bigl( \frac{\sigma^{2}}{G_{0}}+ \frac{2|\Delta^{(1)}_{S}|^{2}}{G_{0}} \Bigr)  + \frac{1}{2} \sum_{n_{y},n_{x'},n_{y'}}\int dy_{0}dy_{1}dx'_{0}dx'_{1}dy'_{0}dy'_{1}  \Bigl(\chi_{S} V^{-1}\chi_{S}+\overline{\Xi_{S}} W^{-1}\Xi_{S} \Bigr) \Bigg]  \nonumber \\
& & -\frac{i}{2}\ln {\rm Det}_{xy}{\cal M}(x,y).
\end{eqnarray}
Here, ${\rm Det}_{xy}$ is the determinant of both $x$ and $y$ spaces in the functional sense.
We assume $\sigma$ and $\Delta^{(1)}_{S}$ are constants, do not depend on spacetime coordinates. 
Hence the effective potential is obtained by its definition $V_{eff} = -\Gamma/(\sum_{n_{x}}\int dx_{0}dx_{1})$:
\begin{eqnarray}
V_{eff} &=&  \frac{\sigma^{2}}{G_{0}} + \frac{2|\Delta^{(1)}_{S}|^{2}}{G_{0}} +\frac{1}{2}\int_{k}\int_{k'} \Bigl( \chi_{S}(k)V^{-1}(k,k')\chi_{S}(k') +  \overline{\Xi_{S}}(k)W^{-1}(k,k')\Xi_{S}(k') \Bigr)  +\frac{i}{2}{\rm tr} \int_{k}   \ln {\rm det}{\cal M},  
\end{eqnarray}
where the determinant becomes
\begin{eqnarray}
{\rm tr} \int_{k}   \ln {\rm det}{\cal M} &=& \int_{k}\ln 
\bigl[k^{2}_{0}-(E^{+}_{+})^{2}\bigr]
\bigl[k^{2}_{0}-(E^{-}_{+})^{2}\bigr]
\bigl[k^{2}_{0}-(E^{+}_{-})^{2}\bigr]
\bigl[k^{2}_{0}-(E^{-}_{-})^{2}\bigr].
\end{eqnarray}
The determinant has totally eight eigenvalues $\pm E^{+}_{+}$, $\pm E^{+}_{-}$, $\pm E^{-}_{+}$ and $\pm E^{-}_{-}$ if there is no degeneracy.
Hereafter, we will take the static approximation, assuming that the order parameters $\chi_{S}$, $\Xi_{S}$ and the non-local interactions $V$ and $W$ have no $k_{0}$- ( energy ) dependence.
This approximation corresponds to the neglection of retardation effects in the interactions.

\vspace{2mm}

Due to the matrix structure of ${\cal M}$, it seems difficult to evaluate ${\rm det}{\cal M}$ with including all types of order parameters.
We will consider several situations: 
\begin{itemize}
\item 
Example (i): $B_{3}\ne 0$, $\sigma\ne 0$, $\chi^{(1)}_{S}\ne 0$, $\overrightarrow{\chi^{(3)}_{S}}\ne 0$ and $\Delta^{(1)}_{S}=\Xi^{(1)}_{S}=\overrightarrow{\Xi^{(3)}_{S}}=0$
( no superconductivity ).
We assume $\overrightarrow{\chi^{(3)}_{S}}=(0,0,\chi^{(3)}_{S3})$.
In this case, the energy eigenvalues become
\begin{eqnarray}
E^{+}_{\pm}(\bmk) &=& \sqrt{v^{2}_{F}k^{2}_{1}+(\sigma+m(k_{2})+\chi^{(1)}_{S}(\bmk)+\chi^{(3)}_{S3}(\bmk))^{2}} \mp\Bigl( v_{F}k_{F}+\mu(k_{2})-\frac{g_{J}\mu_{B}}{2}B_{3}\Bigr),   \\ 
E^{-}_{\pm}(\bmk) &=& \sqrt{v^{2}_{F}k^{2}_{1}+(\sigma+m(k_{2})+\chi^{(1)}_{S}(\bmk)-\chi^{(3)}_{S3}(\bmk))^{2}} \mp\Bigl( v_{F}k_{F}+\mu(k_{2})+\frac{g_{J}\mu_{B}}{2}B_{3}\Bigr).
\end{eqnarray}
By this example, we could examine the magnetic-field induced chiral symmetry breaking ( chiral condensation, CDW/SDW-types ) in our model.
\item 
Example (ii): $\sigma\ne 0$, $\chi^{(1)}_{S} \ne 0$, $\Delta^{(1)}_{S} \ne 0$, $\Xi^{(1)}_{S} \ne 0$, $\overrightarrow{\Xi^{(3)}_{S}}\ne 0$, and $B_{3}=\overrightarrow{\chi^{(3)}_{S}}=0$.
We will choose the spin-triplet Cooper pair as $\overrightarrow{\Xi^{(3)}_{S}}=(0,0,\Xi^{(3)}_{S3})$.
This choice is a special case of the condition for unitary order parameter of spin triplet Cooper pairs $\overrightarrow{\Xi^{(3)}_{S}}\overrightarrow{\overline{\Xi^{(3)}_{S}}}\propto 1$, 
usually be employed in a (3+1)-dimensional theory~[20] and in that case it indicates the time-reversal invariance of a spin-triplet superconducting state.
The energy eigenvalues are obtained as follows:
\begin{eqnarray}
E^{+}_{\pm}(\bmk) &=& \sqrt{\bigl( \sqrt{v^{2}_{F}k^{2}_{1}+[m(k_{2})+\sigma+\chi^{(1)}_{S}(\bmk)]^{2}} \mp [v_{F}k_{F}+\mu(k_{2})] \bigr)^{2} + |\Delta^{(1)}_{S}+\Xi^{(1)}_{S}(\bmk)+\Xi^{(3)}_{S3}(\bmk)|^{2} }, \\
E^{-}_{\pm}(\bmk) &=& \sqrt{\bigl( \sqrt{v^{2}_{F}k^{2}_{1}+[m(k_{2})+\sigma+\chi^{(1)}_{S}(\bmk)]^{2}} \mp [v_{F}k_{F}+\mu(k_{2})] \bigr)^{2} + |\Delta^{(1)}_{S}+\Xi^{(1)}_{S}(\bmk)-\Xi^{(3)}_{S3}(\bmk)|^{2} }.
\end{eqnarray}
$E^{+}_{+}$ and $E^{-}_{+}$ are quasiparticles coming from positive energy states, 
while $E^{+}_{-}$ and $E^{-}_{-}$ are quasiparticles coming from negative energy states.
Due to the coexistence of $\Delta^{(1)}_{S}$, $\Xi^{(1)}_{S}$ and $\Xi^{(3)}_{S3}$,
the (pseudo)spin degeneracies have been resolved,
and totally four branches of the quasiparticle excitation spectra exist in this case. 
When $\Xi^{(3)}_{S3}=0$ in this case, these spectra recover the ordinary BCS-type energy spectra:
\begin{eqnarray}
E_{\pm}(\bmk) &=& E^{+}_{\pm}(\bmk) = E^{-}_{\pm}(\bmk) 
= \sqrt{\bigl( \sqrt{v^{2}_{F}k^{2}_{1}+[m+\sigma+\chi^{(1)}_{S}(\bmk)]^{2}} \mp [v_{F}k_{F}+\mu] \bigr)^{2} + |\Delta^{(1)}_{S}+\Xi^{(1)}_{S}(\bmk)|^{2} }.
\end{eqnarray}
Both $E_{+}$ ( $=E^{+}_{+}=E^{-}_{+}$ ) and $E_{-}$ ( $=E^{+}_{-}=E^{-}_{-}$ ) are doubly degenerate. 
Even if $\chi^{(1)}_{S}=\Delta^{(1)}_{S}+\Xi^{(1)}_{S}=0$,
the upper band $E_{+}(\bmk)$ and the lower band $-E_{-}(\bmk)$ have no intersection when $m+\sigma\ne 0$.
At $\sigma+\chi^{(1)}_{S}(\bmk)+2u\cos k_{2}=0$ with $\Delta^{(1)}_{S}+\Xi^{(1)}_{S}(\bmk)=0$, 
the energy dispersions become $E_{\pm}=v_{F}|k_{1}|\mp (v_{F}k_{F}+2t\cos k_{2})$, and they correspond to $\pm{\cal E}_{\pm}$ discussed in the beginning of this paper.
The pure-spin-triplet superconducting state $\overrightarrow{\Xi^{(3)}_{S}} \ne 0$, $B_{3}=\overrightarrow{\chi^{(3)}_{S}}=\Delta^{(1)}_{S}+\Xi^{(1)}_{S}=0$, 
also gives similar spectra, have two doubly-degenerate excitation branches.
\end{itemize}

\vspace{2mm}

We can construct the case where particles of up-spin component of $\psi$ participate the Cooper pair condensation while particles of down-spin component remain a Fermi gas~[58].
In that case, the projection operator in the isospin space $(1\pm\tau_{3})/2$ should be inserted ( as vertices ) to particle interactions.

\vspace{2mm}

Next, we employ the finite-temperature Matsubara formalism~[59-61].
The $k_{0}$-integration is converted into the Matsubara formalism by the following substitution in our theory:
\begin{eqnarray}
& & \int\frac{dk_{0}}{2\pi i} \to \sum_{n}\frac{1}{\beta}, \quad k_{0} \to i\omega_{n}, \quad \beta \equiv \frac{1}{T}, \quad ( T; {\rm temperature} ), \quad \omega_{n} \equiv \frac{(2n+1)\pi}{\beta}, \quad ( n=0,\pm 1,\pm 2, \cdots ),
\end{eqnarray}
and this method is convenient for our discussin in this paper. 
The four-fermion contact interaction model in (2+1)-dimensions is renormalizable~[50], 
though the renormalization of the effective potential could not be performed in the same way of the usual Lorentz-symmetric case.
In fact, due to the existence of cosine terms in several functions of our theory, 
it seems difficult to employ the dimensional regularization, the standard method of regularization for usual (2+1)-dimensional field theory.
Therefore, we do not consider the renormalization of the theory, and regard ${\cal L}$ as a cutoff model. 
( Integrals of (2+1)-dimensional field theories usually become ultraviolet finite by employing the dimensional regularization. )
As our definition of the gamma matrices, our model is closer to (1+1)-dimensional fermion models than (2+1)-dimensional field theories. 
Due to the Matsubara formalism, we should employ a non-covariant cutoff scheme.
The effective potential $V_{eff}$ is obtained in the following form:
\begin{eqnarray}
& & V_{eff} = \frac{\sigma^{2}}{G_{0}} + \frac{2|\Delta^{(1)}_{S}|^{2}}{G_{0}}   \nonumber \\
& & \quad +\frac{1}{2}\int^{\Lambda}_{-\Lambda}\frac{dk_{1}}{2\pi}\int^{\pi}_{-\pi}\frac{dk_{2}}{2\pi}\int^{\Lambda}_{-\Lambda}\frac{dk'_{1}}{2\pi}\int^{\pi}_{-\pi}\frac{dk'_{2}}{2\pi}
\Bigl( \chi_{S}(\bmk)V^{-1}(\bmk,\bmk')\chi_{S}(\bmk') + \overline{\Xi_{S}}(\bmk)W^{-1}(\bmk,\bmk')\Xi_{S}(\bmk') \Bigr)   \nonumber \\
& & \quad -  \frac{1}{2}\int^{\Lambda}_{-\Lambda}\frac{dk_{1}}{2\pi}\int^{\pi}_{-\pi}\frac{dk_{2}}{2\pi}\Bigl( E^{+}_{+}+E^{-}_{+}+E^{+}_{-}+E^{-}_{-} +\frac{2}{\beta}\ln(1+e^{-\beta E^{+}_{+}})(1+e^{-\beta E^{-}_{+}})(1+e^{-\beta E^{+}_{-}})(1+e^{-\beta E^{-}_{-}}) \Bigr),
\end{eqnarray}
where, the cutoff $\Lambda$ has been introduced to the $k_{1}$-integral.
The regularization of integration is controlled by $\Lambda$ of $k_{1}$-direction due to the asymmetry of the momentum space in our case.
Because we use the linearized dispersion approximation to $k_{1}$-direction to obtain our model from an underlying theory such as a Hubbard-type hopping model of particles on a two-dimensional lattice system, and we only take into account the long-wave-length components of the wave-fields of fermions move toward $x_{1}$-direction,
the relation $\Lambda \gg t,u$ has to be satisfied.
The cutoff $\Lambda$, the transfer $t$ and $u$ have mass dimension $[{\rm mass}]^{1}$, while $v_{F}$ is dimensionless.

\vspace{2mm}

The stationary condition of $V_{eff}$ under the variation with respect to order parameters ( i.e., the gap equations ) will be obtained as follows.
The gap equation for $\sigma$ is found to be
\begin{eqnarray}
0 &=& \frac{\partial V_{eff}}{\partial\sigma}   \nonumber \\
&=& \frac{2\sigma}{G_{0}}-\frac{1}{2}\int^{\Lambda}_{-\Lambda}\frac{dk_{1}}{2\pi}\int^{\pi}_{-\pi}\frac{dk_{2}}{2\pi}\Bigg[ 
\frac{\partial E^{+}_{+}}{\partial\sigma}\tanh\frac{\beta}{2}E^{+}_{+} + \frac{\partial E^{-}_{+}}{\partial\sigma}\tanh\frac{\beta}{2}E^{-}_{+}  +\frac{\partial E^{+}_{-}}{\partial\sigma}\tanh\frac{\beta}{2}E^{+}_{-} + \frac{\partial E^{-}_{-}}{\partial\sigma}\tanh\frac{\beta}{2}E^{-}_{-}
\Bigg],  
\end{eqnarray}
while, for $\chi^{(1)}_{S}(\bmk)$,
\begin{eqnarray}
0 &=& \frac{\partial V_{eff}}{\partial \overline{\chi^{(1)}_{S}}(\bmk)} 
= \frac{1}{2}\int^{\Lambda}_{-\Lambda}\frac{dk'_{1}}{2\pi}\int^{\pi}_{-\pi}\frac{dk'_{2}}{2\pi}V^{-1}(\bmk,\bmk')\chi^{(1)}_{S}(\bmk') -\frac{1}{2}\Bigg[ 
\frac{\partial E^{+}_{+}}{\partial\overline{\chi^{(1)}_{S}}(\bmk)}\tanh\frac{\beta}{2}E^{+}_{+} + \frac{\partial E^{-}_{+}}{\partial\overline{\chi^{(1)}_{S}}(\bmk)}\tanh\frac{\beta}{2}E^{-}_{+} \nonumber \\
& & \qquad \qquad \qquad \qquad \qquad + \frac{\partial E^{+}_{-}}{\partial\overline{\chi^{(1)}_{S}}(\bmk)}\tanh\frac{\beta}{2}E^{+}_{-} + \frac{\partial E^{-}_{-}}{\partial\overline{\chi^{(1)}_{S}}(\bmk)}\tanh\frac{\beta}{2}E^{-}_{-} \Bigg].
\end{eqnarray}
The self-consistent equation for $\chi^{(3)}_{S3}$ will be obtained by the replacement $\partial/\partial\chi^{(1)}_{S}\to\partial/\partial\chi^{(3)}_{S3}$.
On the other hand, the gap equation for the spin-singlet $s$-wave Cooper pair $\Delta^{(1)}_{S}$ is found to be
\begin{eqnarray}
0 &=& \frac{\partial V_{eff}}{\partial\overline{\Delta^{(1)}_{S}}}   \nonumber \\
&=& \frac{2\Delta^{(1)}_{S}}{G_{0}}  -\frac{1}{2}\int^{\Lambda}_{-\Lambda}\frac{dk_{1}}{2\pi}\int^{\pi}_{-\pi}\frac{dk_{2}}{2\pi}\Bigg[ 
\frac{\partial E^{+}_{+}}{\partial\overline{\Delta^{(1)}_{S}}}\tanh\frac{\beta}{2}E^{+}_{+} + \frac{\partial E^{-}_{+}}{\partial\overline{\Delta^{(1)}_{S}}}\tanh\frac{\beta}{2}E^{-}_{+} + \frac{\partial E^{+}_{-}}{\partial\overline{\Delta^{(1)}_{S}}}\tanh\frac{\beta}{2}E^{+}_{-} + \frac{\partial E^{-}_{-}}{\partial\overline{\Delta^{(1)}_{S}}}\tanh\frac{\beta}{2}E^{-}_{-} \Bigg],  \nonumber \\
& & 
\end{eqnarray}
while, for $\Xi^{(1)}_{S}(\bmk)$,
\begin{eqnarray}
0 &=& \frac{\partial V_{eff}}{\partial \overline{\Xi^{(1)}_{S}}(\bmk)} 
= \frac{1}{2}\int^{\Lambda}_{-\Lambda}\frac{dk'_{1}}{2\pi}\int^{\pi}_{-\pi}\frac{dk'_{2}}{2\pi}W^{-1}(\bmk,\bmk')\Xi^{(1)}_{S}(\bmk') -\frac{1}{2}\Bigg[ 
\frac{\partial E^{+}_{+}}{\partial\overline{\Xi^{(1)}_{S}}}\tanh\frac{\beta}{2}E^{+}_{+} + \frac{\partial E^{-}_{+}}{\partial\overline{\Xi^{(1)}_{S}}}\tanh\frac{\beta}{2}E^{-}_{+} \nonumber \\
& & \qquad \qquad \qquad \qquad + \frac{\partial E^{+}_{-}}{\partial\overline{\Xi^{(1)}_{S}}}\tanh\frac{\beta}{2}E^{+}_{-} + \frac{\partial E^{-}_{-}}{\partial\overline{\Xi^{(1)}_{S}}}\tanh\frac{\beta}{2}E^{-}_{-} \Bigg]. 
\end{eqnarray}
The gap equation for the self-consistent determination of $\Xi^{(3)}_{3S}(\bmk)$ will be obtained by replacing $\partial/\partial\Xi^{(1)}_{S}\to\partial/\partial\Xi^{(3)}_{3S}$ in the expression given above.
Several derivatives appeared in the gap equations obtain different forms between the examples (i) and (ii). 
The derivatives appear in example (i) become
\begin{eqnarray}
\frac{\partial E^{+}_{\pm}}{\partial\sigma} &=& \frac{\partial E^{+}_{\pm}}{\partial\chi^{(1)}_{S}} = \frac{\partial E^{+}_{\pm}}{\partial\chi^{(3)}_{S3}} 
= \frac{\sigma+m+\chi^{(1)}_{S}+\chi^{(3)}_{S3}}{\sqrt{v^{2}_{F}k^{2}_{1}+(\sigma+m+\chi^{(1)}_{S}+\chi^{(3)}_{S3})^{2}} },   \\
\frac{\partial E^{-}_{\pm}}{\partial\sigma} &=& \frac{\partial E^{-}_{\pm}}{\partial\chi^{(1)}_{S}} = -\frac{\partial E^{-}_{\pm}}{\partial\chi^{(3)}_{S3}}
= \frac{\sigma+m+\chi^{(1)}_{S}+\chi^{(3)}_{S3}}{\sqrt{v^{2}_{F}k^{2}_{1}+(\sigma+m+\chi^{(1)}_{S}-\chi^{(3)}_{S3})^{2}} }.
\end{eqnarray}
The derivatives with respect to $\sigma$ and $\chi^{(1)}_{S}$ in example (ii) become
\begin{eqnarray}
\frac{\partial E^{+}_{\pm}}{\partial\sigma} &=& \frac{\partial E^{+}_{\pm}}{\partial\chi^{(1)}_{S}}    \nonumber \\
&=& \frac{\sqrt{v^{2}_{F}k^{2}_{1}+[\sigma+\chi^{(1)}_{S}+m]^{2}} \mp [v_{F}k_{F}+\mu]}{\sqrt{\bigl( \sqrt{v^{2}_{F}k^{2}_{1}+[\sigma+\chi^{(1)}_{S}+m]^{2}} \mp [v_{F}k_{F}+\mu] \bigr)^{2} + |\Delta^{(1)}+\Xi^{(1)}_{S}+\Xi^{(3)}_{S3}|^{2}}}\frac{\sigma+\chi^{(1)}_{S}+m}{\sqrt{v^{2}_{F}k^{2}_{1}+(\sigma+m+\chi^{(1)}_{S})^{2}}},   \\
\frac{\partial E^{-}_{\pm}}{\partial\sigma} &=& \frac{\partial E^{-}_{\pm}}{\partial\chi^{(1)}_{S}}   \nonumber \\
&=& \frac{\sqrt{v^{2}_{F}k^{2}_{1}+[\sigma+\chi^{(1)}_{S}+m]^{2}} \mp [v_{F}k_{F}+\mu]}{\sqrt{\bigl( \sqrt{v^{2}_{F}k^{2}_{1}+[\sigma+\chi^{(1)}_{S}+m]^{2}} \mp [v_{F}k_{F}+\mu] \bigr)^{2} + |\Delta^{(1)}_{S}+\Xi^{(1)}_{S}-\Xi^{(3)}_{S3}|^{2}}}\frac{\sigma+\chi^{(1)}_{S}+m}{\sqrt{v^{2}_{F}k^{2}_{1}+(\sigma+m+\chi^{(1)}_{S})^{2}}}.
\end{eqnarray}
The derivatives with respect to $\Delta^{(1)}_{S}$, $\Xi^{(1)}_{S}$ and $\Xi^{(3)}_{S3}$ of the example (ii) become
\begin{eqnarray}
\frac{\partial E^{+}_{\pm}}{\partial\overline{\Delta^{(1)}_{S}}} 
&=& \frac{\partial E^{+}_{\pm}}{\partial\overline{\Xi^{(1)}_{S}}} 
= \frac{\partial E^{+}_{\pm}}{\partial\overline{\Xi^{(3)}_{S3}}}
= \frac{1}{2}\frac{\Delta^{(1)}_{S}+\Xi^{(1)}_{S}(\bmk)+\Xi^{(3)}_{S3}(\bmk)}{\sqrt{\bigl( \sqrt{v^{2}_{F}k^{2}_{1}+[\sigma+\chi^{(1)}_{S}+m]^{2}} \mp [v_{F}k_{F}+\mu] \bigr)^{2} + |\Delta^{(1)}_{S}+\Xi^{(1)}_{S}+\Xi^{(3)}_{S3}|^{2}}},    \\
\frac{\partial E^{-}_{\pm}}{\partial\overline{\Delta^{(1)}_{S}}} 
&=& \frac{\partial E^{+}_{\pm}}{\partial\overline{\Xi^{(1)}_{S}}} 
= -\frac{\partial E^{+}_{\pm}}{\partial\overline{\Xi^{(3)}_{S3}}}
= \frac{1}{2}\frac{\Delta^{(1)}_{S}+\Xi^{(1)}_{S}(\bmk)-\Xi^{(3)}_{S3}(\bmk)}{\sqrt{\bigl( \sqrt{v^{2}_{F}k^{2}_{1}+[\sigma+\chi^{(1)}_{S}+m]^{2}} \mp [v_{F}k_{F}+\mu] \bigr)^{2} + |\Delta^{(1)}_{S}+\Xi^{(1)}_{S}-\Xi^{(3)}_{S3}|^{2}}}.
\end{eqnarray}
The dominant contribution to the integrations of the gap equations of the example (ii) will come from the quasiparticles of the vicinity of $\sqrt{v^{2}_{F}k^{2}_{1}+[\sigma+\chi^{(1)}_{S}+m(k_{2})]^{2}}-[v_{F}k_{F}+\mu(k_{2})]\sim 0$ of the positive energy branch in the $(k_{1},k_{2})$-space.
Therefore, $V_{eff}$ and the gap equations include $G_{0}$, $\Lambda$, $v_{F}$, $u$, $t$, $g_{J}\mu_{B}B_{3}$ and $T$ as external ( model ) parameters.
We will calculate $V_{eff}$ and the gap equations numerically in the space of these model parameters.
In the case of $k_{F}=t=u=\chi^{(1)}_{S}=|\Delta^{(1)}_{S}+\Xi^{(1)}_{S}\pm\Xi^{(3)}_{S3}|=B_{3}=0$ of the example (ii), 
the gap equation for $\sigma$ becomes that of the well-known result of the (1+1)-dimensional Gross-Neveu model by evaluating the method of large-$N$ expansion~[43]:
\begin{eqnarray}
\sigma &=& \sigma \frac{G_{0}}{2}\int^{\Lambda}_{-\Lambda}\frac{dk_{1}}{2\pi}\frac{1}{E(k_{1})}\tanh\frac{\beta}{2}E(k_{1}), \quad E(k_{1}) = E^{\pm}_{\pm}(k_{1},0) =\sqrt{v^{2}_{F}k^{2}_{1}+\sigma^{2}}.
\end{eqnarray}

\vspace{2mm}

Now, we specify the functional forms of $V$ and $W$.
Usually in the generalized BCS framework, the effective attractive interaction will be decomposed into channels by using the complete set of angular momentum eigenfunctions.
In the quasi-(1+1)-dimensional case we consider here, the decompositions of the effective interactions $V(\bmk,\bmk')$ and $W(\bmk,\bmk')$ are simply performed by Fourier expansions in both $k_{1}$ and $k_{2}$ directions. 
Under the weak coupling approximation, $V(\bmk,\bmk')$ can be written in the following form in general:
\begin{eqnarray}
V(\bmk,\bmk') &=& \sum_{n,m}\Bigl( g^{(cc)}_{nm}\cos nk_{1}\cos mk_{2}\cos nk'_{1}\cos mk'_{2} + g^{(cs)}_{nm}\cos nk_{1}\sin mk_{2}\cos nk'_{1}\sin mk'_{2}  \nonumber \\
& & + g^{(sc)}_{nm}\sin nk_{1}\cos mk_{2}\sin nk'_{1}\cos mk'_{2} + g^{(ss)}_{nm}\sin nk_{1}\sin mk_{2}\sin nk'_{1}\sin mk'_{2} \Bigr).
\end{eqnarray}
Here, $|\bmk|=|\bmk'|$ must be satisfied before and after scattering of particles.
Parallel with this procedure, the order parameters $\chi_{S}$ and $\Xi_{S}$ also be decomposed as follows:
\begin{eqnarray} 
\chi_{S} &=& |\chi_{0}| + \sum_{n,m}\Bigl( |\chi^{(cc)}_{nm}|\cos nk_{1}\cos mk_{2} + |\chi^{(cs)}_{nm}|\cos nk_{1}\sin mk_{2}  \nonumber \\ 
& & + |\chi^{(sc)}_{nm}|\sin nk_{1}\cos mk_{2} + |\chi^{(ss)}_{nm}|\sin nk_{1}\sin mk_{2} \Bigr).
\end{eqnarray}
For example, we consider the following components of the partial-wave decompositions:
\begin{eqnarray}
V(\bmk,\bmk') = W(\bmk,\bmk') &\sim& g_{0}, \quad ( s ),\\
&\sim& g_{p_{x}}\sin k_{1}\sin k'_{1},  \quad ( p_{x} ),  \\
&\sim& g_{p_{y}}\sin k_{2}\sin k'_{2},  \quad ( p_{y} ),  \\
&\sim& g_{d_{x^{2}-y^{2}}}\cos k_{2}\cos k'_{2},  \quad ( d_{x^{2}-y^{2}} ),  \\
&\sim& g_{d_{xy}}\sin k_{1}\sin k_{2}\sin k'_{1}\sin k'_{2},  \quad ( d_{xy} ),    \\
&\sim& g_{f_{x}}\sin k_{1}\cos k_{2}\sin k'_{1}\cos k'_{2},  \quad ( f_{x} ),  \\
&\sim& g_{f_{y}}\sin 2k_{2}\sin 2k'_{2},  \quad ( f_{y} ).
\end{eqnarray}
Here, $s$, $p_{x}$, ..., denotes the symmetry of fermion-(anti)fermion condensations, 
and we have identified that $x=k_{1}$ and $y=k_{2}$.
We should notice that, the parity of the order parameters will be determined by both the dependence of $k_{1}$ and $k_{2}$.
Here, we assume $V=W$, while in general, we can choose different interactions between $V$ and $W$.
By putting these several types of interactions in (87)-(93), we will obtain the gap equations under the generalized BCS formalism.

\vspace{2mm}

We briefly comment on the renormalization of $V_{eff}$. 
As mentioned above, due to the anisotropy of the model, the renormalization procedure combined with the dimensional regularization
cannot be performed in our theory. 
Let us consider the case $\Delta=\chi=\Xi=0$ for our example.
In the effective potential evaluated by the cutoff regularization, we can incorporate the following renormalization condition~[43,62]:
\begin{eqnarray}
\frac{\partial^{2}V_{eff}}{\partial \sigma^{2}}\Big|_{\sigma=\mu_{r}} &=& \frac{2}{G_{ren}}.
\end{eqnarray}
Here, $G_{ren}$ is the renormalized coupling constant, $\mu_{r}$ denotes the renormalization point. Therefore, $V_{eff}$ can be written formally as
\begin{eqnarray}
V^{ren}_{eff} &=& \frac{\sigma^{2}}{G_{ren}} -\frac{1}{2}\frac{\partial^{2}I(\sigma)}{\partial\sigma^{2}}\Big|_{\sigma=\mu_{r}}\sigma^{2} + I(\sigma), \quad
I(\sigma) \equiv V_{eff} - \frac{\sigma^{2}}{G_{0}}.
\end{eqnarray}
However, it is unclear for us that whether the cutoff $\Lambda$ is removed or not by introducing the renormalization point in the procedure given above.
At least at the level of analytical formulation, it seems impossible.
Moreover, we have to solve the renormalization group equation for obtaining the results they are independent on $\mu_{r}$. 
When we consider the case $\sigma\ne0$, $\Delta^{(1)}_{S}\ne 0$, $\chi_{S}\ne 0$ and $\Xi_{S}\ne 0$,
the Coleman-Weinberg type renormalization condition will become more complicated one than Eq. (94) ( though we can write it formally ),
and thus, in fact, it it impossible to employ the renormalization procedure for our $V_{eff}$.

\section{Summary}

In this paper we have constructed the BCS and generalized BCS theory for the dynamical mass and superconducting gap generations in the quasi-(1+1)-dimensional relativistic model. We have introduced a "relativistic" ( Dirac ) model where particles move intrachain direction as a homogeneous gas while the motion of particles toward the interchain direction is described by a hopping term similar to the case of Hubbard model. We have performed the group-theoretical consideration on several fermion-(anti)fermion pair functions, and have chosen the suitable order parameters for the dinamical masses and superconducting gaps in the possible situation of condensed matters.
We have constructed the effective potential and the gap equations in analytical manner. 

As the next work, we will present the results of numerical solutions of the gap equations.
The thermodynamic properties, response to external fields will also be given.
Preparation for the presentation of these results is now in progress,
and we will publish them as the part II of this study.


\begin{thebibliography}{999}




\bibitem{kanamori}
J. Kanamori,
{\it Electron Correlation and Ferromagnetism of Transition Metals},
Prog. Theor. Phys. {\bf 30}, 275 (1963).
\bibitem{moriya}
T. Moriya, Y. Takahashi and K. Ueda,
{\it Antiferromagnetic Spin Fluctuations and Superconductivity in Two-Dimensional Metals --- A Possible Model for High $T_{c}$ Oxides},
J. Phys. Soc. Jpn. {\bf 59}, 2905 (1990),
T. Moriya and K. Ueda,
{\it Spin Fluctuation Spectra and High Temperature Superconductivity},
J. Phys. Soc. Jpn. {\bf 63}, 1871 (1994),
T. Takimoto and T. Moriya,
{\it Theory of Spin Fluctuation-Induced Superconductivity Based on a $d-p$ Model},
J. Phys. Soc. Jpn. {\bf 66}, 2459 (1997).
\bibitem{yamaishi}
T. Ishiguro and K. Yamaji,
{\it Organic Conductors},
( Springer, Heidelberg, 1990 ).
\bibitem{kagoshima}
S. Kagoshima, 
{\it Low-Dimensional Conductors: ---Physics of Organic Conductors and Density Waves---}
( Shokabo, Tokyo, 2000 ).
\bibitem{yamaji1}
K. Yamaji,
{\it Semimetallic SDW State in Quasi One-Dimensional Conductors},
J. Phys. Soc. Jpn. {\bf 51}, 2787 (1982).
\bibitem{yamaji2}
K. Yamaji,
{\it First-Order Phase Transition Boundary between Superconducting and SDW Phases in the Bechgaard Salts},
J. Phys. Soc. Jpn. {\bf 52}, 1361 (1983).
\bibitem{imadafujimoritokura}
M. Imada, A. Fujimori and Y. Tokura,
{\it Metal-insulator Transitions},
Rev. Mod. Phys. {\bf 70}, 1039 (1998).
\bibitem{kuroko}
K. Kuroki, R. Arita and H. Aoki,
{\it Spin-Triplet f-wave-like Pairing Proposed for an Organic Superconductor ${\rm (TMTSF)_{2}PF_{6}}$ },
Phys. Rev. {\bf B63}, 094509 (2001),
Y. Tanaka and K. Kuroki,
{\it Microscopic Theory of Spin-Triplet f-wave Pairing in Quasi-One-Dimensional Organic Superconductors},
Phys. Rev. {\bf B70}, 060502(R) (2004),
S. Onari, R. Arita, K. Kuroki and H. Aoki,
{\it Phase Diagram of the Two-Dimensional Extended Hubbard Model: Phase Transitions between Different Pairing Symmetries when Charge and Spin Fluctuations Coexist},
Phys. Rev. {\bf B70}, 094523 (2004).
\bibitem{duprat}
R. Duprat and C. Bourbonnais,
{\it Interplay between Spin-density-wave and Superconducting States in Quasi-one-dimensional Conductors},
Eur. Phys. J. {\bf B21}, 219 (2001).
\bibitem{nickel}
J. C. Nickel, R. Duprat, C. Bourbonnais and N. Dupuis,
{\it Triplet Superconducting Pairing and Density-Wave Instabilities in Organic Conductors},
Phys. Rev. Lett. {\bf 95}, 247001 (2005).
\bibitem{fuseya}
Y. Fuseya and Y. Suzumura,
{\it Superconductivity and Density Wave in the Quasi-One-Dimensional Systems: Renormalization Group Study},
J. Phys. Soc. Jpn. {\bf 74}, 1263 (2005).
\bibitem{supersolid}
A discussion in the $t-J$ model: D. Poilblanc,
{\it Enhanced Pairing in Doped Quantum Magnets with Frustrated Hole Motion},
Phys. Rev. Lett. {\bf 93}, 197204 (2004),
A discussion in the Hubbard model: D. Poilblanc, K. Penc and N. Shannon,
{\it Doped Singlet-pair Crystal in the Hubbard Model on the Checkerboard Lattice},
Phys. Rev. {\bf B75}, 220503(R) (2007).
See also, in a Bose system: P. Sengupta and C. D. Batista,
{\it Field-Induced Supersolid Phase in Spin-One Heisenberg Models},
Phys. Rev. Lett. {\bf 98}, 227201 (2007).
\bibitem{bailin}
D. Bailin and A. Love,
{\it Superfluidity in Ultrarelativistic Quark Matter},
Nucl. Phys. {\bf B190}, 175 (1981),
{\it Superfluidity and Superconductivity in Relativistic Fermion Systems},
Phys. Rep. {\bf B107}, 325 (1984).
\bibitem{colorsuper}
For example:
M. Iwasaki and T. Iwado,
{\it Superconductivity in Quark Matter},
Phys. Lett. {\bf B350}, 163 (1995),
M. Alford, K. Rajagopal and F. Wilczek,
{\it QCD at Finite Baryon Density: Nucleon Droplets and Color Superconductivity},
Phys. Lett. {\it B422}, 247 (1998),
R. Rapp, T. Sch\"{a}fer, E. V. Shuryak and M. Velkovsky,
{\it Diquark Bose Condensates in High Density Matter and Instantons},
Phys. Rev. Lett. {\bf 81}, 53 (1998).
\bibitem{bcs1}
J. Bardeen, L. N. Cooper and J. R. Schrieffer,
{\it Microscopic Theory of Superconductivity},
Phys. Rev. {\bf 106}, 162 (1957),
{\it Theory of Superconductivity},
Phys. Rev. {\bf 108}, 1175 (1957).
L. N. Cooper, 
{\it Bound Electron Pairs in a Degenerate Fermi Gas},
Phys. Rev. {\bf 104}, 1189 (1956).
\bibitem{nambu}
Y. Nambu,
{\it Quasi-particles and Gauge Invariance in the Theory of Superconductivity},
Phys. Rev. {\bf 117}, 648 (1960).
\bibitem{unconv}
P. W. Anderson and P. Morel,
{\it Generalized Bardeen-Cooper-Schrieffer States and the Proposed Low-Temperature Phase of Liquid He3},
Phys. Rev. {\bf 123}, 1911 (1961),
R. Balian and N. R. Werthamer,
{\it Superconductivity with Pairs in a Relative p Wave},
Phys. Rev. {\bf 131}, 1553 (1963).
\bibitem{legett}
A. J. Leggett,
{\it A Theoretical Description of the New Phases of Liquid $^{3}{\rm He}$},
Rev. Mod. phys. {\bf 47}, 331 (1975).
\bibitem{vollwoel}
D. Vollhardt and P. W\"{o}lfle,
{\it The Superfluid Phases of Helium 3},
( Taylor and Francis, London, 1990 ).
\bibitem{tsuneto}
T. Tsuneto,
{\it Superconductivity and Superfluidity},
( Iwanami, Tokyo, 1997 ) ( in Japanese ).
[ English version ( Cambridge University Press, Cambridge, 1998 ), translated by M. Nakahara. ]
\bibitem{nambu2}
Y. Nambu and G. Jona-Lasinio,
{\it Dynamical Model of Elementary Particles Based on an Analogy with Superconductivity. I},
Phys. Rev. {\bf 122}, 345 (1961), 
{\it Dynamical Model of Elementary Particles Based on an Analogy with Superconductivity. II},
Phys. Rev. {\bf 124}, 246 (1961).
\bibitem{ohsaku3}
T. Ohsaku, Thesis, Dept. Phys. Osaka University, Dec. 26, 2000.
\bibitem{ohsaku4}
T. Ohsaku,
{\it BCS and Generalized BCS Superconductivity in Relativistic Quantum Field Theory: Formulation},
Phys. Rev. {\bf B65}, 024512 (2002).
\bibitem{ohsaku5}
T. Ohsaku,
{\it BCS and Generalized BCS Superconductivity in Relativistic Quantum Field Theory. II. Numerical Calculations},
Phys. Rev. {\bf B66}, 054518 (2002).
\bibitem{ohsaku6}
T. Ohsaku,
{\it Relativistic Model of two-band Superconductivity in (2+1)-dimension},
Int. J. Mod. Phys. {\bf B18}, 1771 (2004).
\bibitem{ohsaku7}
T. Ohsaku,
{\it Chiral Symmetry and Collective Excitations in p-wave, d-wave, and f-wave Superconductors},
cond-mat/0306095.
\bibitem{solid}
N. W. Ashcroft and N. D. Mermin,
{\it Solid State Physics},
( Holt, Rinehart and Winston, 1976, New York ).
\bibitem{fradkin}
E. Fradkin, 
{\it Field Theories of Condensed Matter Systems},
( Addison-Wesley, New York, 1991 ).
\bibitem{nagaosa}
N. Nagaosa,
{\it Quantum Field Theory in Condensed Matter Physics},
( Iwanami, Tokyo, 1991 ) ( in Japanese ).
[ English version, ( Springer, Berlin, 1999 ). ];
{\it Quantum Field Theory in Strongly Correlated Electronic Systems},
( Iwanami, Tokyo, 1998 ) ( in Japanese ).
[ English version ( Springer, Berlin, 1999 ). ]
\bibitem{lutting}
W. Kohn and J. M. Luttinger,
{\it New Mechanism for Superconductivity},
Phys. Rev. Lett. {\bf 15}, 524 (1965),
J. M. Luttinger, 
{\it New Mechanism for Superconductivity},
Phys. Rev. {\bf 150}, 202 (1966).
\bibitem{tomo}
S. Tomonaga,
{\it Remarks on Bloch's Method of Sound Waves applied to Many-Fermion Problems},
Prog. Theor. Phys. {\bf 5}, 544 (1950),
J. M. Luttinger,
{\it An Exactly Soluble Model of a Many-Fermion System},
J. Math. Phys. {\bf 4}, 1154 (1963).
\bibitem{thirring}
W. E. Thirring,
{\it A Soluble Relativistic Field Theory},
Ann. Phys. {\bf 3}, 91 (1958).
\bibitem{soltm}
D. C. Mattis and E. H. Lieb,
{\it Exact Solution of Many-Fermion System and Its Associated Boson Field},
J. Math. Phys. {\bf 6}, 304 (1965).
On the G\^{o}to-Imamura-Schwinger term appear in the bosonization, see, 
T. G\^{o}to and T. Imamura,
{\it Note on the Non-Perturbation-Approach to Quantum Field Theory},
Prog. Theor. Phys. {\bf 14}, 396 (1955),
J. Schwinger,
{\it Field Theory Commutators},
Phys. Rev. Lett. {\bf 3}, 296 (1959).
\bibitem{callan}
C. G. Callan, R. F. Dashen and D. H. Sharp,
{\it Solvable Two-Dimensional Field Theory Based on Currents},
Phys. Rev. {\bf 165}, 1883 (1968).
\bibitem{witten}
E. Witten,
{\it Chiral Symmetry, the $1/N$ Expansion and the $SU(N)$ Thirring Model},
Nucl. Phys. {\bf B145}, 110 (1978).
\bibitem{witten2}
E. Witten,
{\it Non-Abelian Bosonization in Two Dimensions},
Commun. Math. Phys. {\bf 92}, 455 (1984).
\bibitem{nonabel}
Y. Frishman and J. Sonnenschein, 
{\it Bosonization and QCD in Two Dimensions},
Phys, Rep. {\bf 223}, 309 (1993).
\bibitem{liebwu}
E. H. Lieb and F. Y. Wu,
{\it Absence of Mott Transition in an Exact Solution of the Short-Range, One-Band Model in One Dimension},
Phys. Rev. Lett. {\bf 20}, 1445 (1968),
M. Takahashi,
{\it Magnetization Curve for the Half-Filled Hubbard Model},
Prog. Theor. Phys. {\bf 42}, 1098 (1969),
{\it On the Exact Ground State Energy of Lieb and Wu},
Prog. Theor. Phys. {\it 45}, 756 (1971).
\bibitem{luther}
A. Luther,
{\it Eigenvalue Spectrum of Interacting Massive Fermions in One Dimension},
Phys. Rev. {\bf B14}, 2153 (1976).
\bibitem{andreilowenstein}
H. Bergknoff and H. B. Thacker,
{\it Method for Solving the Massive Thirring Model},
Phys. Rev. Lett. {\bf 42}, 135 (1079),
{\it Structure and Solution of the Massive Thirring Model},
Phys. Rev. {\bf D19}, 3666 (1979),
N. Andrei and J. H. Lowenstein,
{\it Diagonalization of the Chiral-Invariant Gross-Neveu Hamiltonian},
Phys. Rev. Lett. {\bf 43}, 1698 (1979).
\bibitem{murayamaiso}
S. Iso and H. Murayama,
{\it Hamiltonian Formulation of the Schwinger Model ---Non-Confinement and Screening of the Charge---},
Prog. Theor. Phys. {\bf 84}, 142 (1990).
\bibitem{hosotani}
Y. Hosotani and R. Rodriguez, 
{\it Bosonized N-Flavor Massive Schwinger Model},
J. Phys. {\bf A31}, 9925 (1998),
Y. Hosotani,
{\it Antiferromagnetic $S=1/2$ Heisenberg Chain and the Two-Flavor Masless Schwinger Model},
Phys. Rev. {\bf B60}, 6198 (1999).
\bibitem{grossneveu}
D. J. Gross and A. Neveu,
{\it Dynamical Symmetry Breaking in Asymptotically Free Field Theories},
Phys. Rev. {\bf D10}, 3235 (1974).
\bibitem{mermin}
N. D. Mermin and H. Wagner,
{\it Absence of Ferromagnetism or Antiferromagnetism in One- and Two-Dimensional Isotropic Heisenberg Models},
Phys. Rev. Lett. {\bf 17}, 1133 (1966).
\bibitem{hoenberg}
P. C. Hohenberg,
{\it Existence of Long-Range Order in One and Two Dimensions},
Phys. Rev. {\bf 158}, 383 (1967).
\bibitem{SRColeman}
S. R. Coleman,
{\it There are no Goldstone Bosons in Two-Dimensions},
Commun. Math. Phys. {\bf 31}, 259 (1973).
\bibitem{ktteni}
J. M. Kosterlitz and D. J. Thouless,
{\it Ordering, Metastability and Phase Transitions in Two-dimensional Systems},
J. Phys. {\bf C6}, 1181 (1973).
\bibitem{klimenko}
K. G. Klimenko,
{\it Phase Structure of Generalized Gross-Neveu Models},
Z. Phys. {\bf C37}, 457 (1988).
\bibitem{klimenko2}
K. G. Klimenko,
{\it Three-dimensional Gross-Neveu Model at Nonzero Temperature and in the Presence of an External Electromagnetic Field},
Z. Phys. {\bf B54}, 323 (1992).
\bibitem{fourfermi}
B. Rosenstein, B. J. Warr and S. H. Park,
{\it Four-Fermion Theory is Renormalizable in 2+1 Dimensions},
Phys. Rev. Lett. {\bf 62}, 1433 (1989),
{\it Dynamical Symmetry Breaking in Four-fermion Interaction Models},
Phys. Rep. {\bf 205}, 59 (1991).
\bibitem{inagaki}
T. Inagaki, T. Muta and S. D. Odintsov, 
{\it Dynamical Symmetry Breaking in Curved Spacetime},
Prog. Theor. Phys. Suppl. {\bf 127}, 93 (1997).
\bibitem{ojimafukuda}
I. Ojima and R. Fukuda,
{\it Another Phase in the Gross-Neveu Model. --- Two Dimensional Superconductivity ---},
Yukawa Institute Preprint, RIFP-267, Oct. 1976, Kyoto, Japan.
In this reference, fermion-fermion ( not fermion-antifermion ) pair condensation ( hence, a superconductivity ) was discussed in relativistic quantum field theoretical framework, 
and an important observation on the role of the Pauli-G\"{u}rsey symmetry in the (1+1)-dimensional Gross-Neveu model was given in detail.
( Available at http://www.arxiv.org ).
\bibitem{pauliguersey}
W. Pauli, 
Nuovo Cim. {\bf 6}, 204 (1957),
G. G\"{u}rsey, 
Nuovo Cim. {\bf 7}, 411 (1958).
\bibitem{kahana}
D. Kahana and U. Vogl,
{\it Diquark Bosonization of the Nambu Model},
Phys. Lett. {\bf B244}, 10 (1990).
\bibitem{vafawitten}
C. Vafa and E. Witten,
{\it Restrictions on Symmetry Breaking in Vector-like Gauge Theories},
Nucl. Phys. {\bf B234}, 173 (1984).
\bibitem{kleinert}
H. Kleinert,
{\it Hadronization of Quark Theories and a Bilocal Form of QED},
Phys. Lett. {\bf 62B}, 429 (1976).
\bibitem{kugo}
T. Kugo,
{\it Dynamical Instability of the Vacuum in the Lagrangian Formalism of the Bethe-Salpeter Bound States},
Phys. Lett. {\bf 76B}, 625 (1978).
Several papers in literature formulated the bilocal auxiliary fields for composites. 
This reference discussed a bilocal auxiliary field which is described by the center of mass coordinate.
Hence, this is appropriate for describing a Cooper pair, 
because it is at least convenient for us to choose the origin of the Fermi sphere as the center of mass of Cooper pairs. 
\bibitem{kitaoka}
A. Harada, S. Kawasaki, H. Mukuda, Y. Kitaoka, Y. Haga, E. Yamamoto, Y. \={O}nuki, K. M. Itoh, E. E. Haller and H. Harima,
{\it Experimental Evidence for Ferromagnetic Spin-Pairing Superconductivity Emerging in ${\rm UGe}_{2}$: A $^{73}{\rm Ge}$-Nuclear-Quadrupole-Resonance Study under Pressure},
Phys. Rev. {\bf B75}, 140502(R) (2007).
\bibitem{matsu}
T. Matsubara,
{\it A New Approach to Quantum-Statistical Mechanics},
Prog. Theor. Phys. {\bf 14}, 351 (1955).
\bibitem{etu}
H. Ezawa, Y. Tomozawa and H. Umezawa,
{\it Quantum Statistics of Fields and Multiple Production of Mesons},
Nuovo Cim. {\bf 5}, 810 (1957).
\bibitem{finite}
A. A. Abrikosov, L. P. Gor'kov and I. E. Dzyaloshinskii,
{\it Methods of Quantum Field Theory in Statistical Physics},
( Dover, New York, 1963 ),
A. L. Fetter and D. J. Walecka, 
{\it Quantum Theory of Many-particle Systems},
( McGraw-Hill, New York, 1971 ),
J. I. Kapusta, 
{\it Finite-temperature Field Theory},
( Cambridege University Press, Cambridge, 1989 ).
\bibitem{colewein}
S. R. Coleman and E. Weinberg,
{\it Radiative Corrections as the Origin of Spontaneous Symmetry Breaking},
Phys. Rev. {\bf D7}, 1888 (1973).
%\bibitem{zirnbauer}
%M. R. Zirnbauer,
%{\it Supersymmetry for Systems with Unitary Disorder: Circular Ensembles},
%J. Phys. {\bf A29}, 7113 (1996).



\end{thebibliography}
\end{document}